\documentclass[iop]{emulateapj}
\usepackage{amsmath}
\usepackage{graphicx}
\usepackage{rotating}
\usepackage{subfloat}
\usepackage{color}
\usepackage{natbib}
\usepackage{soul}
\usepackage{subfigure}
\usepackage{threeparttable}
\setcitestyle{authoryear,round,semicolon,aysep={,},yysep={,},notesep={; }}
\begin{document}
\title{ High-precision photometric redshifts from Spitzer/IRAC: EXTREME [3.6]-[4.5] COLORS identify GALAXIES IN THE REDSHIFT RANGE $\lowercase{z}\sim6.6-6.9$}

\author{Renske Smit\altaffilmark{1,2}, Rychard J. Bouwens\altaffilmark{1}, Marijn Franx\altaffilmark{1}, Pascal A. Oesch\altaffilmark{3}, Matthew L. N. Ashby\altaffilmark{4}, S. P. Willner\altaffilmark{4}, Ivo Labb\'{e}\altaffilmark{1}, Benne Holwerda\altaffilmark{1}, Giovanni G. Fazio\altaffilmark{4}, J.-S. Huang\altaffilmark{4}}

\altaffiltext{1}{Leiden Observatory, Leiden University, NL-2300 RA Leiden, Netherlands} 
\altaffiltext{2}{Centre for Extragalactic Astronomy, Durham University, South Road, Durham, DH1 3LE, UK} 
\altaffiltext{3}{Yale Center for Astronomy and Astrophysics, Yale University, New Haven, CT 06520, USA} 
\altaffiltext{4}{Harvard-Smithsonian Center for Astrophysics, 60 Garden St., Cambridge, MA 02138, USA}

\begin{abstract}
One of the most challenging aspects of studying galaxies in the $z\gtrsim7$ universe is the infrequent confirmation of their redshifts through spectroscopy, a phenomenon thought to occur from the increasing opacity of the intergalactic medium to Ly$\alpha$ photons at $z>6.5$. 
The resulting redshift uncertainties inhibit the efficient search for [\ion{C}{2}] in $z\sim7$ galaxies with sub-mm instruments such as ALMA, given their limited scan speed for faint lines.
One means by which to improve the precision of the inferred redshifts is to exploit the potential impact of strong nebular emission lines on the colors of $z\sim4-8$ galaxies as observed by \textit{Spitzer}/IRAC. At $z\sim6.8$, galaxies exhibit IRAC colors as blue as $[3.6]-[4.5]\sim-1$, likely due to the contribution of [\ion{O}{3}]+H$\beta$ to the 3.6$\,\micron$ flux combined with the absence of line contamination in the 4.5$\,\micron$ band. In this paper we explore the use of extremely blue $[3.6]-[4.5]$ colors to identify galaxies in the narrow redshift window $z\sim6.6-6.9$. When combined with an $I$-dropout criterion, we demonstrate that we can plausibly select a relatively clean sample of $z\sim6.8$ galaxies.
Through a systematic application of this selection technique to our catalogs from all five CANDELS fields, we identify 20 probable $z\sim6.6-6.9$ galaxies.  We estimate that our criteria select the $\sim50$\% strongest line emitters at $z\sim6.8$ and from the IRAC colors we estimate a typical [\ion{O}{3}]$\rm +H\beta$ rest-frame equivalent width of 1085{$\,$\AA} for this sample. The small redshift uncertainties on our sample make it particularly well suited for follow-up studies with facilities such as ALMA.
\end{abstract}

\keywords{Galaxies: high-redshift --- Galaxies: evolution}
\section{Introduction}
\label{sec:intro}

Since the installation of the Wide Field Camera 3 (WFC3) on the \textit{Hubble Space Telescope (HST)}, numerous ultraviolet (UV) bright galaxies in the reionization era have been detected through their broadband photometric properties. These observations allow for the determination of the UV luminosity function \citep[e.g.,][]{Bouwens2011,Lorenzoni2011,Oesch2012,Oesch2013,Oesch2014,Bradley2012,Bowler2012,Bowler2014,Schenker2013a,Mclure2013} and the typical UV colors of galaxies out to $z\sim8$ \citep{Stanway2005,Bouwens2009, Bouwens2012,Bouwens2013,Dunlop2012,Dunlop2013,Wilkins2011,Finkelstein2012}. 

In contrast to the success of identifying candidate galaxies out to $z\sim8$ with \textit{HST}/WFC3 imaging, confirming the redshift of these sources with spectroscopy has proven very challenging due to the absorption of Ly$\alpha$ photons by the neutral Intergalactic Medium (IGM) at $z\gtrsim6.5$ \citep{Pentericci2011, Treu2013,  Finkelstein2013,Caruana2013, Schenker2014}. This creates particular challenges for follow-up studies with the newest generation of sub-mm telescopes such as ALMA. 
ALMA has the potential to perform detailed studies of star formation rates (SFRs), kinematic structure, and energetics of $z\sim6-8$ galaxies through the direct detection of sub-mm fine structure lines such as [\ion{C}{2}]$\lambda$157.7$\,\micron$ \citep[e.g.][]{Carilli2013}. However, the frequency range that ALMA is able to scan in one tuning is relatively small, making follow-up studies with ALMA on sources without accurate redshift information observationally expensive.  

One way to make progress in this area involves a search for the \ion{C}{3}]$\lambda$1908{$\,$\AA}  line in $z\gtrsim6$ galaxies \citep{Stark2014a,Stark2014b}. This line has a typical rest-frame equivalent width (EW$_0$) of $\sim4-14${$\,$\AA} in two $z\sim6-7$ galaxies where this line has been successfully located \citep{Stark2014b} and in low-mass lensed star-forming galaxies at $z\sim2$ \citep{Stark2014a}.   The challenge with this approach is the faintness of the \ion{C}{3}] line and high density of sky lines in many regions of the near-infrared (IR) spectrum.

Another potentially promising way forward is to utilize the information provided by the \textit{Spitzer}/IRAC.
 Recent studies have reported evidence for the presence of strong nebular emission lines such as H$\alpha$ and [\ion{O}{3}]$\lambda$5007{$\,$\AA} through the apparent impact of these lines on the IRAC 3.6$\,\micron$ and 4.5$\,\micron$ fluxes of $z\sim4-8$ galaxies \citep{Schaerer2009,Shim2011,Stark2013,Labbe2013,Gonzalez2012,Gonzalez2014,Smit2014}. These lines appear to cause the $[3.6]-[4.5]$ colors of high redshift galaxies to vary significantly as a function of redshift. This results in modestly blue $[3.6]-[4.5]$ colors in $z\sim6$ galaxies, where both bands are contaminated by emission lines, very blue $[3.6]-[4.5]$ colors at $z\sim6.8$, where only the 3.6$\,\micron$ band suffers line contamination, and red $[3.6]-[4.5]$ colors for sources at redshifts $z>7$, where the 4.5$\,\micron$ band is contaminated \citep{Labbe2013,Wilkins2013,Smit2014,Laporte2014}.
    
In this paper we attempt to exploit the extreme IRAC colors galaxies exhibit at specific redshifts to isolate galaxies over a narrow range in redshift. To test this method, we utilize a large sample of $z\sim5-8$ galaxies identified from the CANDELS program \citep{Grogin2011,Koekemoer2011}. We select a sample of sources with extremely blue $[3.6]-[4.5]$ IRAC colors and show that these sources likely fall in the narrow redshift range $z\sim6.6-6.9$. Ultimately, of course, this approach and the assumptions behind it will need to be tested through spectroscopy with the \textit{James Webb Space Telescope (JWST)}. 

Additionally we discuss a few  objects with blue $[3.6]-[4.5]$ colors that are clearly at $z\lesssim6.6$ from their \textit{HST} photometry, and we suggest that these objects could be explained by  high [\ion{O}{3}]/H$\beta$ ratios such as found at $z\sim3$ \citep{Schenker2013b,Holden2014}.

This paper is organized as follows. \S2 discusses our data set and photometric procedure. \S3 presents the $[3.6]-[4.5]$ IRAC colors as a function of photometric redshift for a sample of  $z\sim5-8$ UV-selected galaxies. \S4 discusses the selection of extreme $[3.6]-[4.5]$ color galaxies, and \S5 presents our $z\sim6.6-6.9$ galaxy sample. \S6 gives a short summary of our results.

Throughout this paper we adopt a Salpeter IMF with limits 0.1-100$\,M_\odot$ \citep{Salpeter}. For ease of comparison with previous studies we take $H_0=70\,\rm km\,s^{-1}\,Mpc^{-1},\,\Omega_{\rm{m}}=0.3,\,$and$\,\Omega_\Lambda=0.7$. Magnitudes are quoted in the AB system \citep{OkeGun}

\begin{figure}
\centering
\includegraphics[width=0.8\columnwidth,trim=30mm 15mm 80mm 10mm] {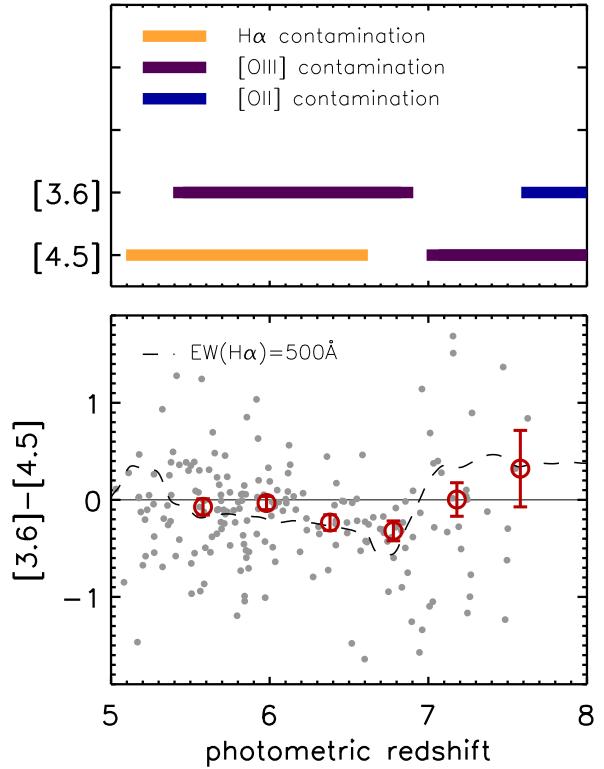} 
\caption{\textit{Top panel:} A schematic overview of the redshift range where the strongest nebular lines (H$\alpha$, [\ion{O}{3}] and [\ion{O}{2}]) can contaminate the rest-frame optical flux in the 3.6$\,\micron$ and 4.5$\,\micron$ bands. 
\textit{Bottom panel:} Measurements of the $[3.6]-[4.5]$ color (grey points) in UV-selected galaxies from GOODS-N/S, placed at their photometric redshift. The red points indicate the median colors in $\Delta z=0.4$ redshift bins (error bars represent the uncertainty in the median, i.e., $\sigma/\sqrt{N}$). The black dashed line provides an example of the IRAC color at different redshifts for a source with flat continuum and EW$_0(\rm H\alpha)=500${$\,$\AA} in combination with emission line ratios as defined by \citet{Anders2003} for sub-solar metallicity $\rm Z=0.2Z_\odot$. The overall blue $[3.6]-[4.5]$ colors at $z<7$, versus red $[3.6]-[4.5]$ colors at $z>7$ indicate that in particular the presence of [\ion{O}{3}] in the IRAC bands might have a significant impact on the $[3.6]-[4.5]$ color. }
\label{fig:emlines_median}
\end{figure}

\section{Observations, photometry and $\lowercase{z}\sim5-8$ sample}
\label{sec:Observations}

\subsection{\textit{HST} data and photometry}
\label{sec:obs_hst}

The primary sample of $z\sim5-8$ galaxies that we use in this paper was selected from the catalogs described by \citet{Bouwens2014}. The purpose of this sample is to establish how the $[3.6]-[4.5]$ color of galaxies depends on redshift and also to establish the redshift distribution of galaxies with the most extreme IRAC colors.  The catalogs were compiled from \textit{HST} ACS ($B_{435},V_{606},i_{775},I_{814}$ and $z_{850}$) and WFC3/IR ($Y_{098}/Y_{105},J_{125},JH_{140}$ and $H_{160}$) data over the GOODS-N and GOODS-S fields.  We also used the deep and wide-area observations obtained in the HUDF09+HUDF12 \citep{Beckwith2006,Bouwens2011,Ellis2013,Illingworth2013}, ERS \citep{Windhorst2011} and CANDELS \citep{Grogin2011,Koekemoer2011} programs as well as any archival \textit{HST} observations over these fields. The shallow $JH_{140}$ imaging was taken from the 3D-\textit{HST} survey \citep{Brammer2012} and A Grism H-Alpha Spectroscopic survey (AGHAST, PI:Weiner). The procedure for reducing the data is described in detail by \citet{Illingworth2013} and \citet{Bouwens2014}.

A secondary sample of $z\sim5-8$ galaxies is used  to increase the number of $z\sim6.8$ and $z\sim6.0$ galaxies in our selection. We derived this sample from the \textit{HST} data over the CANDELS-UDS, CANDELS-COSMOS, and CANDELS-EGS fields  ($V_{606},I_{814},J_{125}$ and $H_{160}$, for more details on the CANDELS fields see \citealt{Koekemoer2011} and \citealt{Skelton2014}) and deep $U$- and $B$-band ground-based observations from CFHT and Subaru \citep{Capak2007,Furusawa2008}. 

An overview of all fields, bands and depths is given in Table 1 of \citet{Bouwens2014}. Both the ACS and WFC3/IR data reach total magnitudes of $\gtrsim27.2$ at 5$\sigma$, using as a basis the flux uncertainties on the total magnitude measurements for the faintest 20\% of galaxies from the \citet{Bouwens2014} catalogs over these fields.   Our total search area over all the fields is 720 arcmin$^2$. 

The photometry followed the procedure described by \citet{Bouwens2012}.  In short, we ran an adapted version of the Source Extractor software \citep{Bertin1996} in dual-image mode.  The detection images were created by combining all deep bands redwards of the Lyman break  (i.e., $Y_{098}/Y_{105},\,J_{125},$ and $H_{160}$) into a square-root $\chi^2$ image \citep{Szalay1999}.  After matching the observations to the $H_{160}$-band point spread function (PSF), colors and total magnitudes were measured in Kron-like apertures with Kron factors 1.6 and 2.5 respectively (defined on the $H_{160}$-band).

\subsection{\textit{Spitzer}/IRAC data and photometry}
\label{sec:obs_irac}

The first part of our \textit{Spitzer}/IRAC data set covers all five CANDELS fields with the 3.6$\,\micron$ and 4.5$\,\micron$ bands\footnote{http://irsa.ipac.caltech.edu/data/SPITZER/docs/ irac/calibrationfiles/spectralresponse/} from the \textit{Spitzer} Extended Deep Survey (SEDS, PI: Fazio) and all available archival data sets from before 2011 \citep{Ashby2013}.  We complement this data set with the new deep IRAC imaging from the \textit{Spitzer} Very Deep Survey (S-CANDELS) Exploration Science Project (Ashby et al., submitted), which brings the IRAC coverage up to 50 hours in depth (26.8 mag at 5$\sigma$ in the 3.6$\,\micron$ band). For the sources in the HUDF, we utilize additional data from the IRAC Ultra Deep Field \citep[IUDF:][]{Labbe2013} program. 

	Before performing photometry on the sources in our sample, we removed the contamination of foreground sources with an automated cleaning procedure \citep{Labbe2010a,Labbe2010b}.  In short, the \textit{HST} images provided a high-spatial resolution template with which to model the positions and flux profiles of the foreground sources.  The light profiles of individual sources in the \textit{HST} image were convolved with a kernel to match the IRAC PSF and then simultaneously fitted to the IRAC image within a region of $\sim11$ arcseconds around the sources from our sample. We subtracted the flux from the foreground galaxies and performed photometry in 2\farcs0-diameter circular apertures on the resulting images. We applied a correction to account for the flux outside of the aperture, given by the ratio of the flux enclosed in the photometric aperture in the \textit{HST} image (before convolution) to the IRAC model (after convolution). This correction ranges from $\sim2.2\times$ to $\sim2.4\times$, depending on the size of the source. The local noise was estimated from the clean background on a residual IRAC image from which all sources were subtracted.  Our procedure for deblending can fail when contaminating sources are too bright or too close to the central source. We removed sources from our sample with a  high $\chi^2$ parameter determined from the residual IRAC image (see \S\ref{sec:selection}).

\subsection{Base sample of $z\sim5-8$ galaxies}
\label{sec:selection}
In this section, we present the base sample of $z\sim5-8$ galaxies we use to study how the IRAC colors of star-forming galaxies depend upon redshift.
We selected our sources from which we measure IRAC colors in the rest-frame UV, adopting the Lyman-break technique \citep{Steidel1999} with the requirement that the source drops out in the $I_{814}$ band. Specifically, our requirements for $z\sim5-8$ Lyman Break Galaxies (LBGs) were 
\[(I_{814}-J_{125}>0.8)\,\wedge\,(J_{125}-H_{160}<0.4)\]
\begin{equation}
\wedge\,(I_{814}-J_{125}>2(J_{125}-H_{160})+0.8),
\end{equation}
where $\wedge$ indicates logical AND. We chose the $I_{814}$ and $J_{125}$  band fluxes for our identification of $z\gtrsim5$ Lyman Break galaxies, instead of more closely spaced passbands like $z_{850}$ and $Y_{105}$, in order to select galaxies over a relatively extended redshift range $z\sim5.5$ to $z\sim8.5$ (without any gaps) and to use filters that are available over all five CANDELS fields for a more uniform selection.    Information from the more closely spaced bandpasses was nevertheless utilised in deriving photometric redshifts for sources from our selection and hence determining the relationship between the IRAC colors of sources and their redshifts.

We also required sources to have either a non-detection in the $V_{606}$ band ($<2\sigma$) or to have a very strong Lyman break, i.e., $V_{606}-J_{125}>2.5$.  
Furthermore, we required sources to be undetected ($<2\sigma$) in the available $B_{435}$-band data over GOODS North and South or, in the case of the CANDELS UDS, COSMOS, or EGS fields, to be undetected ($<2.5\sigma$) in the $\chi^2$ statistic image \citet{Bouwens2014} derived from the ground-based $U$ and $B$ images. 
We required the SExtractor stellarity parameter (equal to 0 and 1 for extended and point sources, respectively) in the $ J_{125}$ band be less than 0.92  to ensure that our selection is largely free of contamination by stars \citep[e.g.][]{Holwerda2014}. Moreover, the blue IRAC color criterion introduced in \S\ref{sec:selection} also selects effectively against contamination from brown dwarfs in our Milky Way (\citealt{Kirkpatrick2011},Holwerda et al., in prep).  We selected sources with a signal-to-nose ratio $S/N(H_{160})>5$, and additionally we required $f_\nu( H_{160})/ef_\nu(\rm 3.6\,\micron)>2.5$ and $f_\nu( H_{160})/ef_\nu(\rm 4.5\,\micron)>2.5$, where  $f_\nu( H_{160})$ is the measured flux density in the $H_{160}$ band and $ef_\nu(\rm 3.6\,\micron)$ and $ef_\nu(\rm 4.5\,\micron)$ are the estimated uncertainties in the 3.6$\,\micron$ and 4.5$\,\micron$ band fluxes. Our chosen requirements explicitly do not include a dependence on the observed flux in the IRAC bands to ensure that our results are not biased according to the emission line flux in our candidates. In practice, most high-redshift sources exhibit a somewhat red UV-to-optical color \citep{Gonzalez2012}, and as a result 75\% of our sample is detected at $>5\sigma$ in the 3.6$\,\micron$ IRAC band. We exclude those $\sim30$\% of the sources that show strong residuals in the IRAC images after our deblending procedure (\S\ref{sec:obs_irac}), which results in a final sample of 220 sources in GOODS-N/S and 224 sources in CANDELS-EGS/UDS/COSMOS. 

\section{$[3.6]-[4.5]$ color vs. redshift}
Before moving onto a discussion of how the IRAC $[3.6]-[4.5]$ color might be used to refine redshift determinations for specific $z>6$ selections,
it is useful to quickly assess whether our sample agrees with our main assumption: that the $[3.6]-[4.5]$ color is strongly influenced by the presence of strong nebular emission lines such as H$\alpha$ and [\ion{O}{3}] in the IRAC filters. To this end we explore the variation of the median $[3.6]-[4.5]$ color as a function of redshift for our sample. 

Figure \ref{fig:emlines_median} shows the colors of our GOODS-N/S sample as a function of the photometric redshift, as derived from the \textit{HST} broadband photometry. We used the software Easy and Accurate Zphot from Yale \citep[EAZY;][]{Brammer2008} to estimate photometric redshifts for the galaxies in our sample. We used the standard template set of EAZY, but we complemented these templates with a number of templates generated with Galaxy Evolutionary Synthesis Models \citep[GALEV;][]{Kotulla2009}, which includes nebular continuum and emission lines as described by \citet{Anders2003}. Additionally we included a template with an [\ion{O}{3}]$\lambda$4959,5007{$\,$\AA}/H$\beta$ ratio of 10 to match the most extreme line ratios observed in spectroscopy of galaxies at lower redshifts \citep{Amorin2012,Brammer2012b,Jaskot2013,Schenker2013b,vanderwel2013,Holden2014,Steidel2014}.
No use of the \textit{Spitzer}/IRAC photometry is made in the photometric redshift determination to avoid coupling between our redshift estimates and the measured IRAC fluxes. We do not include sources from our EGS/UDS/COSMOS sample in Figure \ref{fig:emlines_median} because of the lack of deep \textit{HST} coverage in the $z_{850}$ and $Y_{105}/Y_{098}$ bands, which is needed for obtaining sufficiently accurate redshifts for the analysis we describe. 

\begin{figure}
\centering
\includegraphics[width=0.8\columnwidth,trim=70mm 35mm 50mm 20mm] {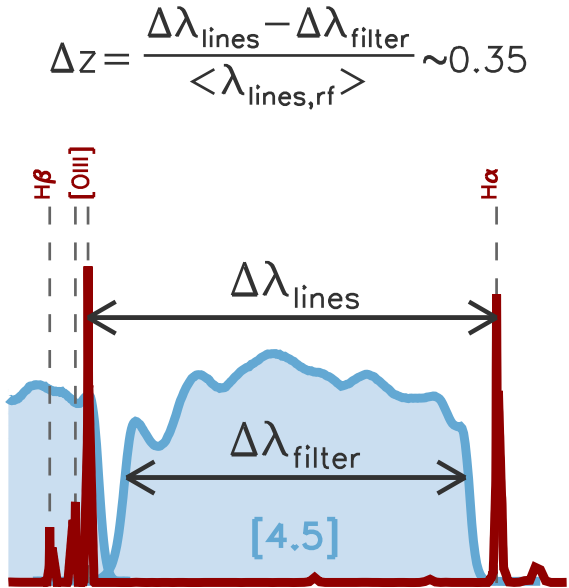} 
\caption{An illustration of the position of the optical nebular emission lines in a $z\sim6.8$ star-forming galaxy (red line) with respect to the  \textit{Spitzer}/IRAC response function (indicated in blue). Our strategy for selecting star-forming galaxies in the narrow redshift range $z\sim6.6-6.9$ capitalizes on the [\ion{O}{3}] and H$\alpha$ emission lines being separated by almost the same  wavelength difference as the width of \textit{Spitzer}/IRAC 4.5$\,\micron$ band.  
Our $[3.6]-[4.5]$ selection technique furthermore requires that the [\ion{O}{3}] line has not yet dropped out of the 3.6$\,\micron$ band, which narrows the redshift window where we expect to find these ultra-blue galaxies, from $\Delta z\sim0.35$ to $\Delta z\sim0.25$.  }
\label{fig:illustration}
\end{figure}

\begin{figure}
\centering
\includegraphics[width=0.8\columnwidth,trim=30mm 0mm 83mm 45mm] {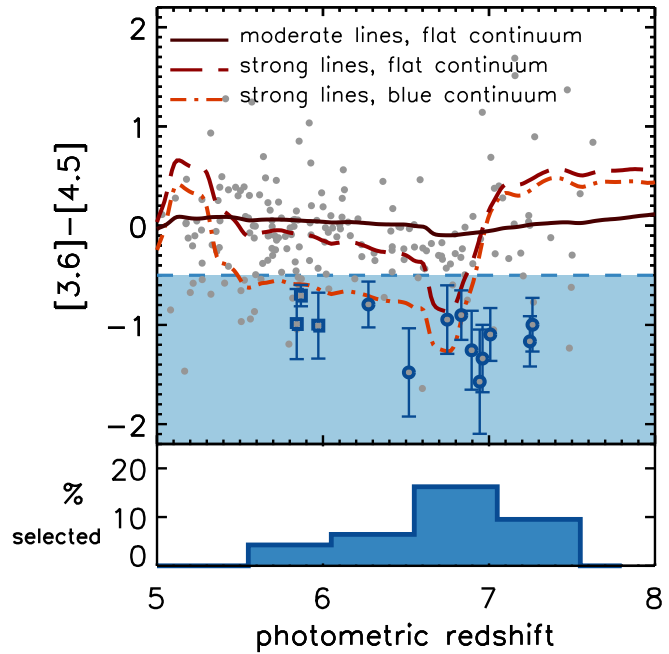} 
\caption{\textit{Top panel:}  Measurements of the $[3.6]-[4.5]$ color (grey points) in UV-selected galaxies from GOODS-N/S, placed at their photometric redshift (see also bottom panel of Figure \ref{fig:emlines_median}). The blue encircled points indicate sources that show IRAC colors significantly bluer than $[3.6]-[4.5]<-0.5$ (blue shaded area), as given by Eq. \ref{eq:blue} (the ultra-blue $[3.6]-[4.5]$ colors for many sources in the blue shaded area are not significant). Blue circles indicate sources that are consistent with the redshift range $z\sim6.6-6.9$, while blue squares indicate sources that are at $z<6.6$ from their photometric redshift probability distribution (99\% confidence). The solid lines indicate three tracks of galaxy templates: the dark red solid line indicates a stellar population with moderate emission lines ($\rm EW_{H\alpha,0}\sim100\,$\AA) and a flat continuum, the red dashed lines indicates a young ($\sim$5 Myr) stellar population with strong emission lines  ($\rm EW_{H\alpha,0}\sim1000\,$\AA) and a flat continuum from moderate dust extinction ($\rm E(B-V)\sim0.2$), while the dot-dashed orange line indicates a young population with strong emission lines, in combination with a high [\ion{O}{3}]/H$\beta$ ratio such as described in \citet{Anders2003} for low metallicity gas ($Z=0.2Z_\odot$) and with blue continuum (dust-free). 
\textit{Bottom panel:} The percentage of the galaxy population that has blue $[3.6]-[4.5]$ colors, such as defined by Eq. \ref{eq:blue}, at a given photometric redshift. The presence of extremely blue $[3.6]-[4.5]$ colors is most abundant in galaxy candidates at $z\sim7$.}
\label{fig:emlines_selection}
\end{figure}

\begin{figure*}[h]
\centering
\includegraphics[width=0.8\textwidth,trim=0mm -5mm 0mm 0mm] {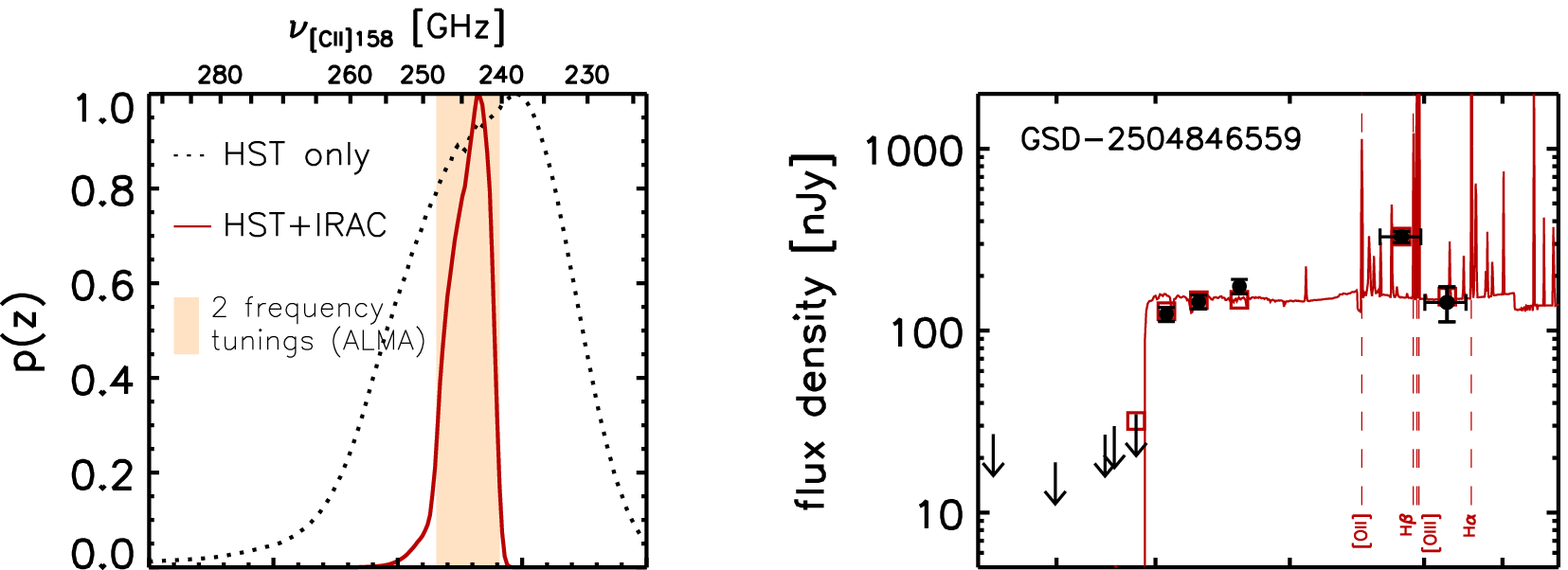} 
\includegraphics[width=0.8\textwidth,trim=0mm -5mm 0mm 0mm] {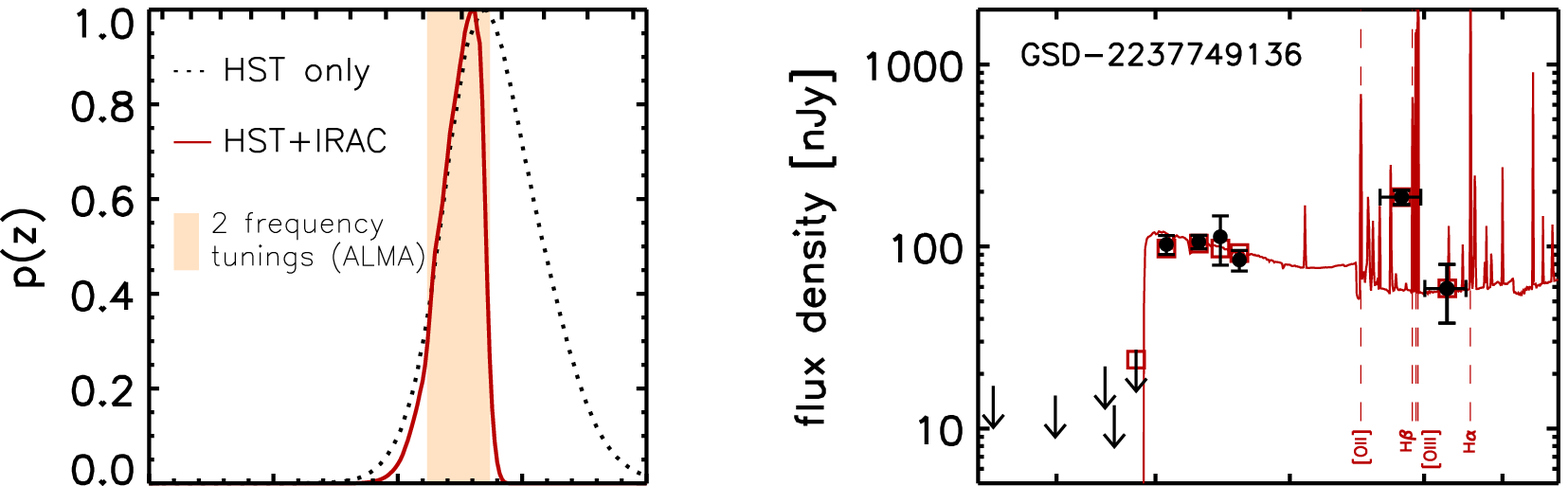} 
\includegraphics[width=0.8\textwidth,trim=0mm -5mm 0mm 0mm] {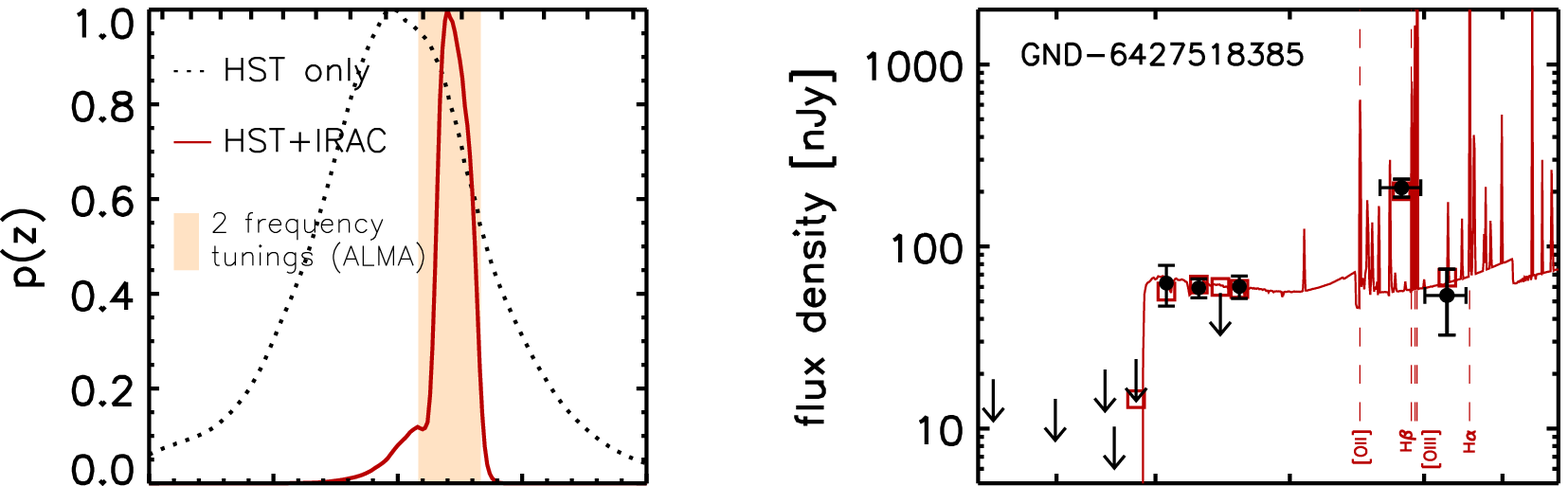} 
\includegraphics[width=0.8\textwidth,trim=0mm -5mm 0mm 0mm] {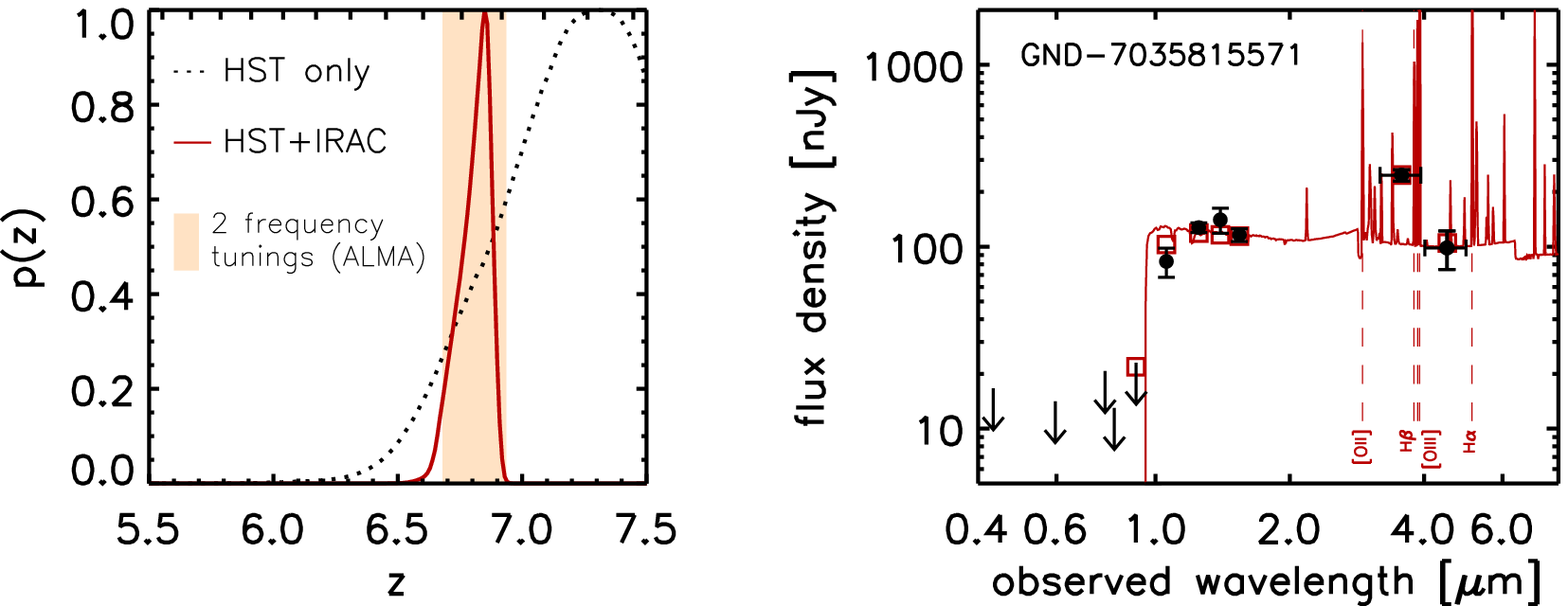} 
\caption{ Four examples of sources selected on their blue IRAC colors (Eq. \ref{eq:blue}) that have a photometric redshift probability distribution consistent (within the 95\% confidence interval) with the redshift range $z\sim6.6-6.9$; from top to bottom GSD-2504846559, GSD-2237749136, GND-6427518385, and GND-7035815571 (from the larger \citealt{Bouwens2014} catalogs). \textit{Left panels:} the redshift probability distribution using only the \textit{HST} bands (black dotted line) and using the constraints from both \textit{HST} and \textit{Spitzer}/IRAC (red line). Due to the substantial impact [\ion{O}{3}] emission can have on the 3.6$\,\micron$ flux, the $[3.6]-[4.5]$ color can set tight constraints on the redshifts of individual sources. Two ALMA tunings (indicated with the yellow shaded regions) would be sufficient to obtain a spectroscopic redshift through the [\ion{C}{2}]$\lambda$158$\,\micron$ line if one is present within the $\sim95$\% likelihood window. \textit{Right panel:} Flux densities and 2$\sigma$ upper limits (black points and arrows) of the \textit{HST} and \textit{Spitzer}/IRAC photometry with the best-fit template (red line). For sources in the redshift window $z\sim6.6-6.9$, the 4.5$\,\micron$ band is not contaminated by strong emission lines, in contrast with the 3.6$\,\micron$ band which is dominated by the flux in the [\ion{O}{3}] line. } 
\label{fig:examples_z6.8}
\end{figure*}

We observe a clear discontinuity in the median color (red points) around $z\sim7$, where the [\ion{O}{3}] emission line moves from the 3.6$\,\micron$ to the 4.5$\,\micron$ band. 
Moreover, we find the bluest median $[3.6]-[4.5]$ color at $z\sim6.8$, where [\ion{O}{3}] boosts the 3.6$\,\micron$ flux while H$\alpha$ has already moved out of the 4.5$\,\micron$ band. This suggests that the IRAC colors are strongly influenced by emission lines, in agreement with recent studies \citep{Schaerer2009,Shim2011,Stark2013,Labbe2013,Gonzalez2012,Gonzalez2014,Smit2014} and in agreement with predictions of stellar population synthesis models with emission lines. We can use this information to improve our determinations of the redshift probability functions for strong line emitters by including the IRAC fluxes in our photometric redshift estimates.

\section{Ultra-blue $[3.6]-[4.5]$ galaxies}
In the previous section we showed how the $[3.6]-[4.5]$ IRAC color would likely change as the [\ion{O}{3}] and H$\alpha$ nebular lines move in and out of the photometric bands, due to the redshifting spectrum. We can use this information to significantly improve our photometric redshift estimates. This is important due to the considerable challenges involved in improving redshift estimates through spectroscopy (largely due to the impact of the more neutral IGM on the prevalence of Ly$\alpha$ emission in $z>6.5$ galaxies).

\subsection{Blue $[3.6]-[4.5]$ sources at $z\sim6.8$}
\label{sec:z6.8}
A particularly interesting redshift interval is $z\sim6.6-6.9$, where we expect extreme $[3.6]-[4.5]$ colors due to the presence of the [\ion{O}{3}] line in the 3.6$\,\micron$ band and the absence of any strong emission lines in the 4.5$\,\micron$ band. This is illustrated in Figure \ref{fig:illustration}. Because the H$\alpha$ to [\ion{O}{3}] line separation ($\Delta\lambda_{\rm lines}$) at $z=6.8$ is slightly wider than the 4.5$\,\micron$ filter width ($\Delta\lambda_{\rm filter}$), there is very narrow redshift range $\Delta z=\left(\Delta\lambda_{\rm lines}-\Delta\lambda_{\rm filter}\right)/\left<\lambda_{\rm lines,rest-frame}\right>=0.35$ in which the 4.5$\,\micron$ band is free of strong emission lines. 
In practice, the effective redshift range where we can observe the extremely blue IRAC colors is even narrower than this, i.e., $\Delta z \sim 0.25$, because the relevant wavelength is where [\ion{O}{3}] leaves the 3.6$\,\micron$ filter, not where this line enters the 4.5$\,\micron$ filter (see the top panel of Figure \ref{fig:emlines_median}).

To investigate whether the selection of sources with blue $[3.6]-[4.5]$ colors can be used to unambiguously identify galaxies at $z\sim6.8$, we collected a sample of emission line candidate galaxies by selecting sources with $[3.6]-[4.5]$ colors significantly bluer than $-0.5$ mag. We adopted the criterion 
\begin{equation}
\label{eq:blue}
P([3.6]-[4.5] < -0.5) > 84\%.
\end{equation}
This criterion indicates that the sources in our selection have a probability (P) of at least 84\% to have a true IRAC color bluer than $[3.6]-[4.5]< -0.5$, based on the uncertainties in the 3.6$\,\micron$ and 4.5$\,\micron$ fluxes.  We show the galaxies meeting this color criterion in Figure \ref{fig:emlines_selection}.  A color cut in the $[3.6]-[4.5]$ color can identify galaxies with strong emission lines in the redshift range $z\sim6.6-6.9$, but a galaxy with more moderate emission lines and very blue continuum $[3.6]-[4.5]$ colors can also meet the criterion. 

We purposely decided to try to select $z\sim6.8$ galaxies based on a color criterion rather than fitting emission line templates to the IRAC bands to obtain photometric redshifts. 
This was done to avoid a dependence on the SED template set and the assumed line ratios in these templates.
Since the physical properties of the \ion{H}{2} regions in star-forming galaxies, such as gas metallicity and gas density, strongly influence the line ratios we observe \citep[e.g.][]{Kewley2013} we wanted to avoid having our results implicitly depend on these ratios.

 Over the 270 arcmin$^2$ CANDELS/ERS region of GOODS- North and GOODS-South, 13 sources satisfy this criterion out of the 220 sources from our base $z\sim5-8$ sample.  We indicate these in the top panel of Figure \ref{fig:emlines_selection} and show a histogram in the bottom panel of Figure \ref{fig:emlines_selection}. Furthermore there are 15 sources out of 224 sources from the base sample over a 450 arcmin$^2$ area in the CANDELS-EGS/CANDELS-UDS/CANDELS-COSMOS fields that meet the criterion, but we do not present them in Figure \ref{fig:emlines_selection} due to the greater difficulty in determining their photometric redshifts.

We identify a large number of blue $[3.6]-[4.5]$ sources that are broadly consistent with a $z\sim6.8$, similar to the sources found by \citet{Smit2014}. For these galaxies the extreme $[3.6]-[4.5]$ colors are explained by the likely scenario that the [\ion{O}{3}] emission dominates the 3.6$\,\micron$ flux while at the same time the 4.5$\,\micron$ flux is free of emission line contamination. 
Interestingly enough, Figure \ref{fig:emlines_selection} indicates there might be a few sources at slightly lower redshift (at $z\sim6.0$ instead of at $z\sim6.8$) with such blue IRAC colors; we will discuss these sources in \S\ref{sec:z6}. 

Figure \ref{fig:examples_z6.8} shows four examples of blue $[3.6]-[4.5]$ sources in GOODS-N/S that are consistent with a $z\sim6.8$ from their \textit{HST} photometry only. We present the redshift probability function using only the \textit{HST} bands and also when using constraints from both \textit{Spitzer}/IRAC and \textit{HST}. Due to their extreme $[3.6]-[4.5]$ colors we can place very tight constraints on the photometric redshift of these galaxies. 

This is particularly useful for follow-up studies to obtain line detections with sub-mm detectors such as ALMA. 
Line detections with interferometer arrays are inherently limited in frequency space by the capabilities of the correlator. 
ALMA can roughly observe $\sim4$ GHz in one tuning in band 6 (211-275 GHz)\footnote{A. Lundgren, 2013, ALMA Cycle 2 Technical Handbook Version 1.1, ALMA, ISBN 978-3-923524-66-2}. As a reference we indicate the frequency of the bright [\ion{C}{2}]$\lambda$157.7$\,\micron$ line at a given redshift in Figure \ref{fig:examples_z6.8} and we indicate the frequency range that ALMA can observe in two frequency tunings. With only twice the observing time required with respect to spectroscopically confirmed sources, we can typically search the $\sim95$\% probability window for [\ion{C}{2}] emission in these sources.

The selection of galaxies at $z\sim6.8$ is also of interest for deriving stellar masses of high-redshift galaxies using \textit{Spitzer}/IRAC constraints \citep{Eyles2005,Eyles2007,Yan2005,Yan2006,Labbe2006,Labbe2010b,Yabe2009,Stark2009,Gonzalez2010,Gonzalez2011,CurtisLake2013}. In particular, the redshift range $z\sim6.6-7.0$ provides us with the only opportunity beyond $z\gtrsim5.4$ to measure the rest-frame optical stellar continuum without the contamination of nebular line emission in the 4.5$\,\micron$ band, allowing for more accurate stellar mass and specific star formation rate estimates \citep[e.g.,][]{Smit2014}.

\subsection{Blue $[3.6]-[4.5]$ sources at $z\sim6.0$}
\label{sec:z6}
In the previous section we showed that a large number of sources with ultra-blue $[3.6]-[4.5]$ colors very likely have redshifts in the narrow range $6.6-6.9$, where the blue colors can be easily explained by the presence of [\ion{O}{3}] in the 3.6$\,\micron$ band, in contrast to the 4.5$\,\micron$ band that contains no strong line emission. 

Figure \ref{fig:emlines_selection} also shows a number of very blue $[3.6]-[4.5]$ sources that prefer a redshift around $z\sim6.0$ and that have a probability of less than $<1\%$ of being at $z\sim6.6-6.9$ (based on the photometric redshift probability distribution using only \textit{HST} bands). In the redshift range $z\sim5.4-6.6$, [\ion{O}{3}] still contaminates the 3.6$\,\micron$ band, but the strong H$\alpha$ line also boosts the 4.5$\,\micron$ flux. Both lines are expected to be strong in young, actively star-forming galaxies, and therefore it is unclear what the explanation is for the significant spread in $[3.6]-[4.5]$ colors and in particular the very blue $[3.6]-[4.5]$ colors observed for a small fraction of the population.  While galaxies containing Active Galactic Nuclei (AGN) can exhibit extremely high [{\ion{O}{3}]/H$\beta$ ratios, this phenomenon is rare in local galaxies with stellar masses below $M_\ast<10^{10}M_\odot$ \citep[e.g.,][]{Juneau2013}. For our sample of star-forming galaxies at $z\sim6.0$, with UV-luminosities ranging from $M_{\rm UV}\sim{-19}$ to $M_{\rm UV}\sim{-21}$, we expect nearly all sources to have stellar masses below this limit \citep[e.g.,][]{Gonzalez2011,Salmon2014}. However, we cannot completely rule out this option based on the limited photometric information available for these sources.

\begin{figure}
\centering
\includegraphics[width=0.88\columnwidth,trim=0mm -5mm 0mm 0mm] {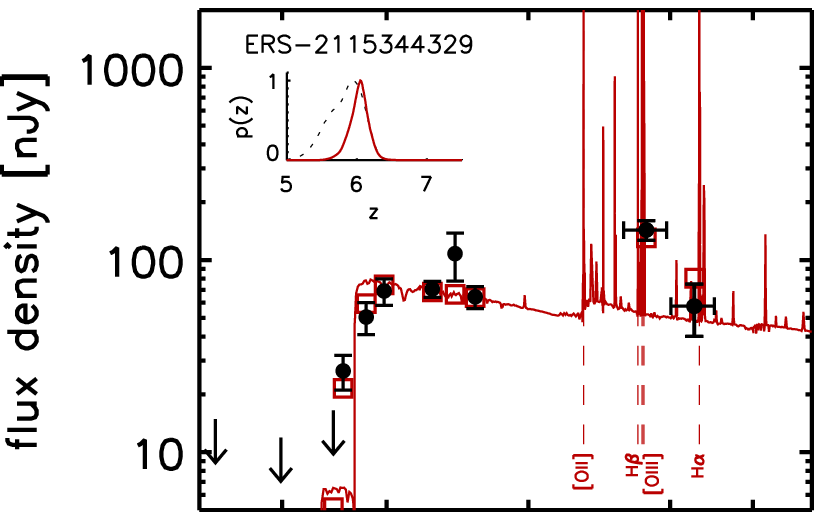} 
\includegraphics[width=0.88\columnwidth,trim=0mm -5mm 0mm 0mm] {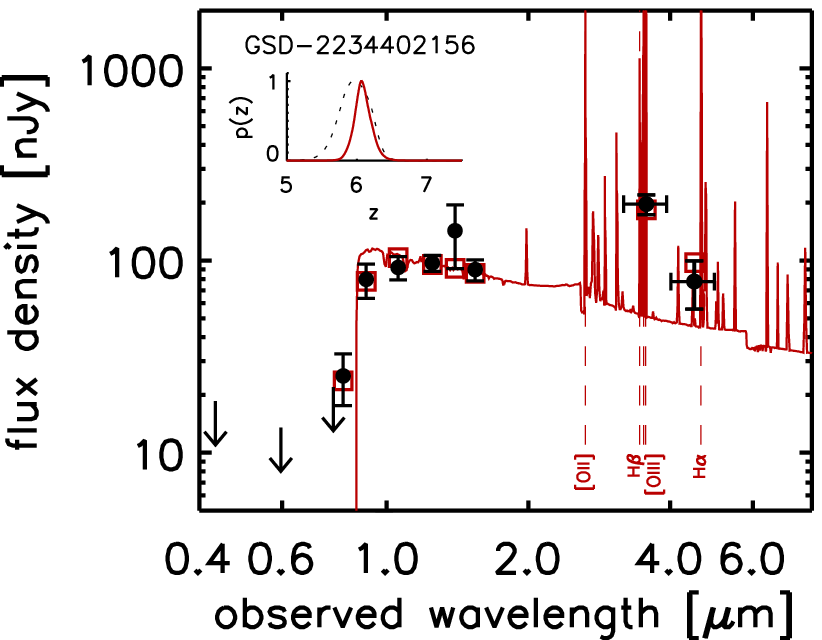} 
\caption{ Two examples of sources that prefer a redshift below $z<6.6$ (ERS-2115344329 and GSD-2234402156) from their photometric redshift distribution (99\% confidence), but that also satisfy our $[3.6]-[4.5]$ color criterion (Eq. \ref{eq:blue}). Flux densities and upper limits (2$\sigma$) of the \textit{HST} and \textit{Spitzer}/IRAC photometry are indicated with black points and arrows, while the best-fit template is drawn in red. The 4.5$\,\micron$ band is contaminated by H$\alpha$, while the 3.6$\,\micron$ band is dominated by the flux in the [\ion{O}{3}] line.  
The inset panel in the top left corner indicates the redshift probability distribution using only the \textit{HST} bands (black dotted line) and using the constraints from both \textit{HST} and \textit{Spitzer}/IRAC (red line).} 
\label{fig:examples_z6}
\end{figure}

\begin{figure}[h]
\centering
\includegraphics[width=0.88\columnwidth,trim=100mm 20mm 10mm 10mm] {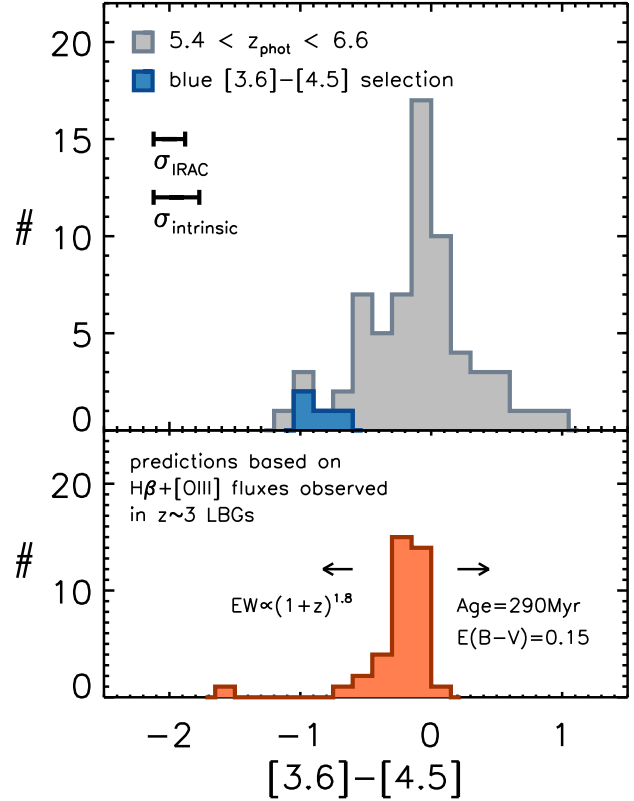} 
\caption{ \textit{Top panel:} The $[3.6]-[4.5]$ color distribution (grey histogram) of sources with photometric redshifts (from the \textit{HST} photometry) within the redshift range $z\sim5.4-6.6$ (68\% confidence) where [\ion{O}{3}] contributes to the 3.6$\,\micron$ flux, while H$\alpha$ contributes to the 4.5$\,\micron$ flux. Sources that satisfy our ultra-blue IRAC selection criterion (Eq. \ref{eq:blue}), but prefer a $z<6.6$ are indicated with the blue histogram (note that for a few sources the very blue $[3.6]-[4.5]$ colors are not significant). The upper error bar on the left side of the panel indicates the scatter in the $[3.6]-[4.5]$ color distribution due to photometric uncertainty in the IRAC bands. The lower error bar indicates the intrinsic scatter, as calculated from the observed scatter and the simulated IRAC uncertainties. \textit{Bottom panel:} The predicted $[3.6]-[4.5]$ color distribution (orange histogram) using the [\ion{O}{3}] and H$\beta$ EWs as measured by \citet{Schenker2013b} and \citet{Holden2014} in $z\sim3$ LBGs with the MOSFIRE spectrograph. We predict the $[3.6]-[4.5]$ color assuming Case B recombination and a flat continuum in $f_\nu$. Evolution of the EW strength as a function of redshift \citep[e.g.,][]{Smit2014} will broaden the $[3.6]-[4.5]$ color distribution and at the same time shift the median color of the distribution to bluer values as indicated by the left black arrow on the panel. The right black arrow indicates the shift in $[3.6]-[4.5]$ color when we assume a somewhat more evolved stellar population with an age of 290 Myr and a dust content of $\rm E(B-V)=0.15$ (see \S\ref{sec:z6} for details).  }  
\label{fig:hist}
\end{figure}

Figure \ref{fig:examples_z6} shows two examples of sources that are at $z<6.6$ at high confidence. One source has an IRAC color of $[3.6]-[4.5]=-1.0 \pm 0.4$ but a photometric redshift of $5.84^{+0.25}_{-0.30} $; the other source has an IRAC color of $[3.6]-[4.5]=-1.0 \pm 0.3$ but a photometric redshift of $5.97^{+0.22}_{-0.22} $.  In both cases, we clearly cannot use the IRAC colors alone to distinguish between these sources and the sources from Figure \ref{fig:examples_z6.8}, which strongly prefer a redshift around $z\sim6.8$. 

One effect that could influence our estimated photometric redshifts for these objects -- and possibly offer an explanation for their blue $[3.6]-[4.5]$ colors -- is  the presence of high EW Ly$\alpha$  emission in these galaxies. 
For example \citet{Schenker2014} show that a  160{\AA} EW Ly$\alpha$ line can cause the photometric redshift of a $z\sim7.5$ galaxy to be underestimated by as much as $\Delta z=0.2$, suggesting that our blue $z\sim6.0$ sample may actually be at higher redshift.   
Though our templates do not include strong Ly$\alpha$ lines, searches for Ly$\alpha$ have shown a low abundance of Ly$\alpha$ emission  at $z>6.5$  \citep{Pentericci2011, Treu2013,  Finkelstein2013, Caruana2013,Schenker2014}, and even at lower redshifts high-EW Ly$\alpha$ is rarely detected in UV bright ($M_{\rm UV}<-20$) galaxies \citep[e.g.][]{Stark2010}. Therefore it seems unlikely to suppose that these blue sources with $z_{phot}\sim6.0$ are actually $z\sim6.8$ galaxies with strong Ly$\alpha$ emission.

In order to understand the origin of these extreme IRAC colors at $z\sim6.0$ we compare the $[3.6]-[4.5]$ color distribution of these sources  with the colors we would predict based on a sample of $z\sim3$ galaxies with near-IR spectroscopy in Figure \ref{fig:hist}. In the top panel of Figure \ref{fig:hist} we show the sub-sample of sources that (with 68\% confidence) have a redshift in the range $z\sim5.4-6.6$, where [\ion{O}{3}] emission contaminates the 3.6$\,\micron$ band while at the same time H$\alpha$ emission contaminates the 4.5$\,\micron$ band (see the top panel of Figure \ref{fig:emlines_median}).  In the bottom panel of Figure \ref{fig:hist} we show a prediction of the $[3.6]-[4.5]$ color distribution from the spectroscopic properties of [\ion{O}{3}] and H$\beta$ in $z\sim3$ LBGs as listed by \citet{Schenker2013b} and \citet{Holden2014}. These authors find line ratios as high as [\ion{O}{3}]$\lambda$4959,5007{$\,$\AA}/H$\beta\sim10$. This strong [\ion{O}{3}] flux with respect to the hydrogen Balmer lines could result in very blue $[3.6]-[4.5]$ colors if similar sources are present at $z\sim6$. We computed the predicted $[3.6]-[4.5]$ colors from the observed [\ion{O}{3}]$\lambda$4959,5007{$\,$\AA} EW and an estimate of the  H$\alpha$ EW derived from the H$\beta$ EW, assuming case B recombination and a flat continuum in $f_\nu$. For $([3.6]-[4.5])_{\rm continuum}=0$ and assuming all sources are at $z=6$, the [\ion{O}{3}] and H$\beta$ EWs allow for a direct calculation of the $[3.6]-[4.5]$ colors such as shown in the bottom panel of Figure \ref{fig:hist}.

\begin{figure}
\centering
\includegraphics[width=0.8\columnwidth,trim=100mm 45mm 10mm 10mm] {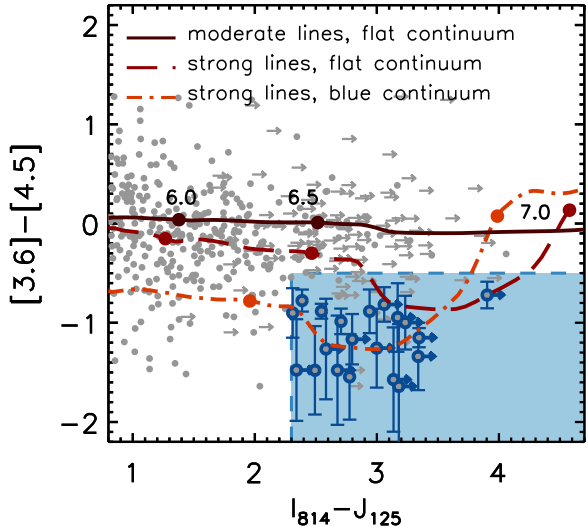} 
\caption{Color-color diagram showing the selection criteria for our fiducial $z\sim6.8$ sample (blue shaded region) with the color measurements of galaxies from our base as sample of  $z\sim5-8$ galaxies over all five CANDELS fields (grey points and arrows; non-detections in the $I_{814}$ band are placed at the 1$\sigma$ upper limit). The blue encircled points indicate the 20 selected sources listed in Table \ref{tab:prop}, that satisfy our blue IRAC criterion (Eq. \ref{eq:blue}) as well as a $I_{814}-J_{125}>2.3$ (see \S\ref{sec:final_sample}) criterion (Sources that have large uncertainties in the IRAC color are not selected). The solid lines indicate tracks for three different galaxy templates as described in Figure \ref{fig:emlines_selection}, with the solid points indicating the colors of the templates at $z=[6.0,6.5,7.0]$. }
\label{fig:selection}
\end{figure}

The spread in colors in the predicted distribution is smaller than the spread in the observed distribution at $z\sim6.0$. Even when accounting for the spread in $[3.6]-[4.5]$ due to the observational uncertainties in the IRAC bands, the intrinsic spread exceeds the width of the predicted distribution (see the error bars in the top panel of Figure \ref{fig:hist}). However, this is likely explained by the spread in the color of the underlying continuum emission due to different dust content and ages of the galaxies in the observed distribution, while we assumed a fixed flat continuum in $f_\nu$ for our predicted colors. A somewhat evolved population ($\sim$300 Myr) and modest dust content ($\rm E(B-V)\sim0.15$) could redden the IRAC color of some of the observed galaxies by $\Delta ([3.6]-[4.5])\sim0.2$ (see Figure \ref{fig:hist}). Similarly, the $[3.6]-[4.5]$ color of the continuum can be as blue as $-0.4$ mag for a very young ($\lesssim10$Myr) and dust-free galaxy.

\begin{figure*}[ht]
\centering
  \subfigure{\includegraphics[width=.45\columnwidth]{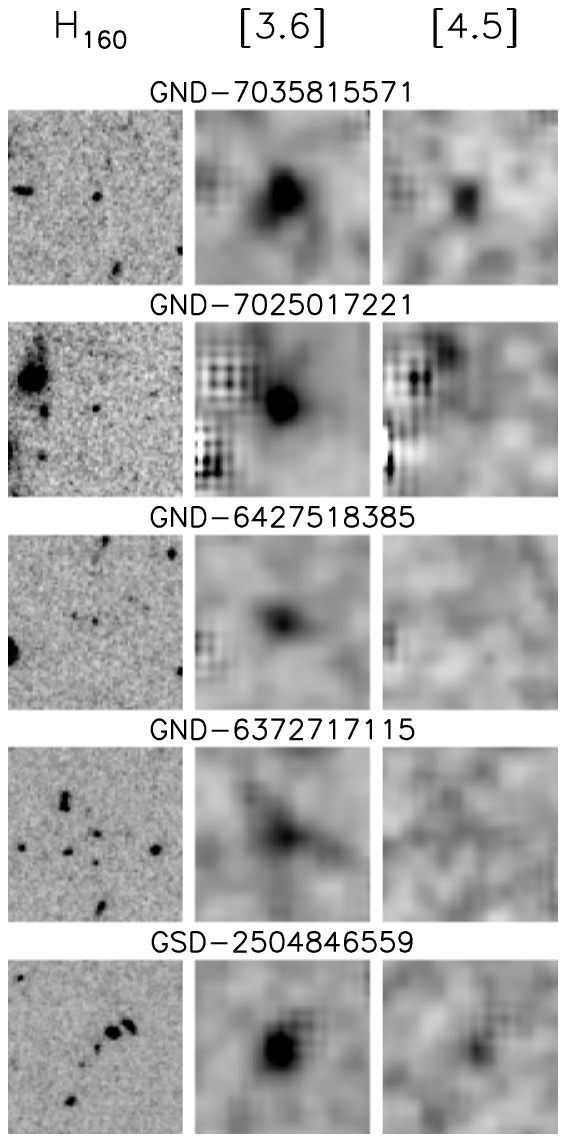}}\quad
  \subfigure{\includegraphics[width=.45\columnwidth]{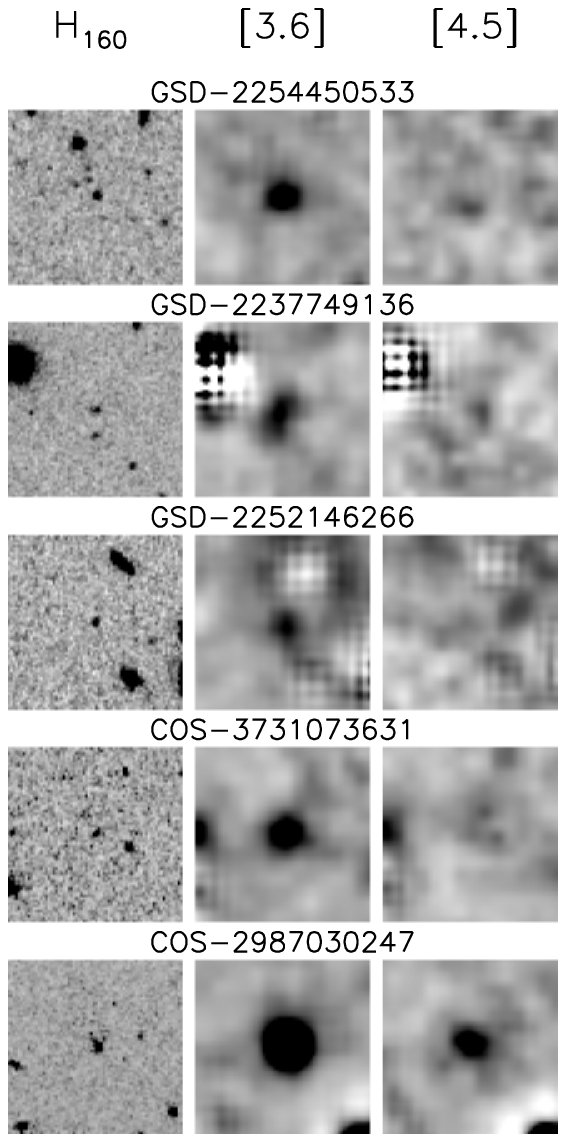}}\quad
  \subfigure{\includegraphics[width=.45\columnwidth]{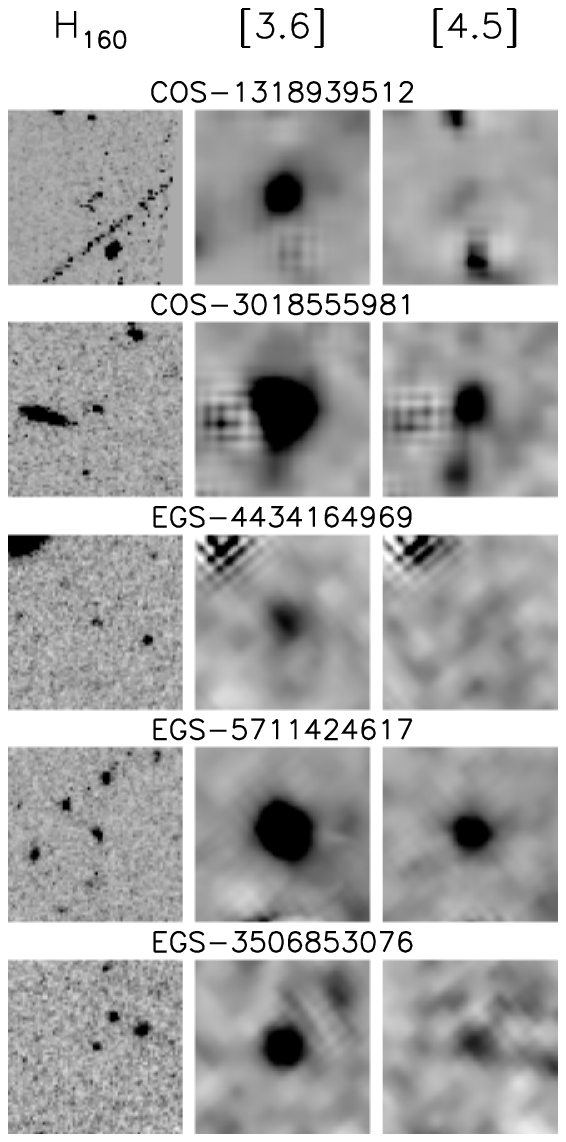}}\quad
  \subfigure{\includegraphics[width=.45\columnwidth]{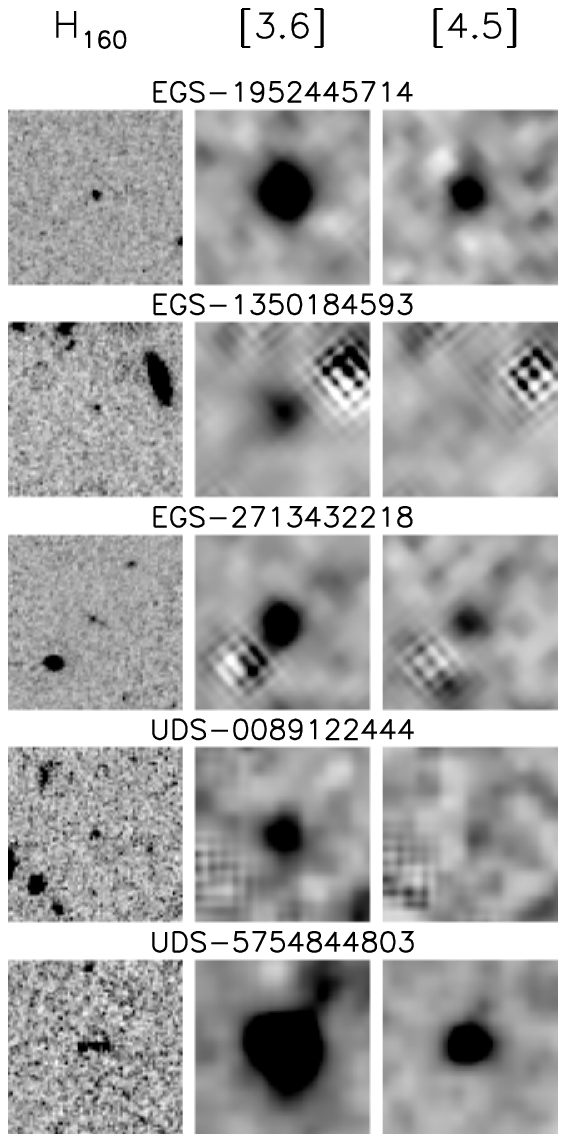}}
\caption{\textit{HST} $H_{160}$, \textit{Spitzer}/IRAC 3.6$\,\micron$, and 4.5$\,\micron$ band postage stamp negative images (8\farcs4 $\times$ 8\farcs4) of our fiducial sample of  galaxies at $z\sim6.8$ in the CANDELS fields with extremely blue $[3.6]-[4.5]$ IRAC colors (satisfying Eq. \ref{eq:blue}) and $I_{814}-J_{125}>2.3$ (see \S\ref{sec:final_sample}). The IRAC postage stamps have been cleaned for contamination from neighboring sources (see \S\ref{sec:obs_irac}). Properties of the sources are listed in Table \ref{tab:prop}.}
\label{fig:stamps}
\end{figure*}

\begin{figure}[h]
\centering
  \subfigure{\includegraphics[width=.45\columnwidth]{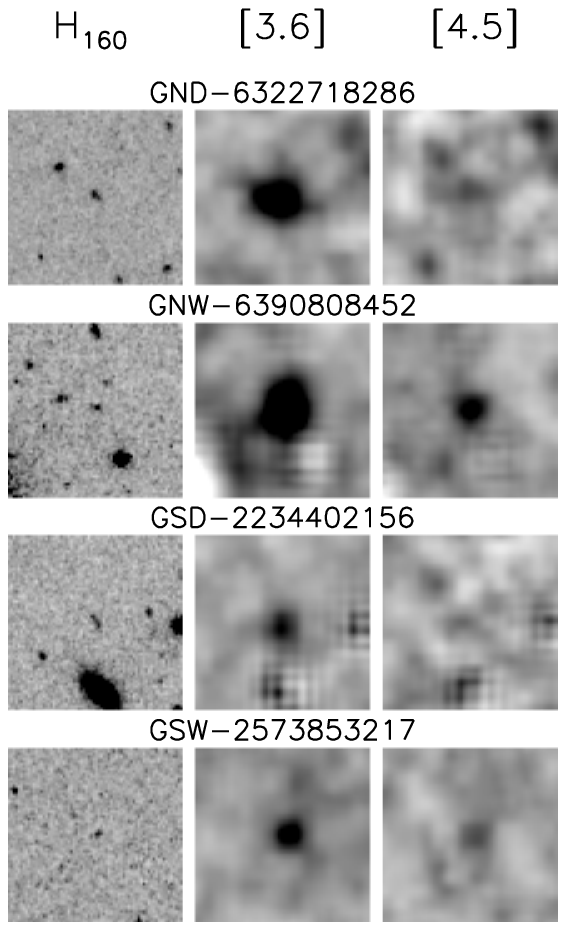}}\quad
  \subfigure{\includegraphics[width=.45\columnwidth]{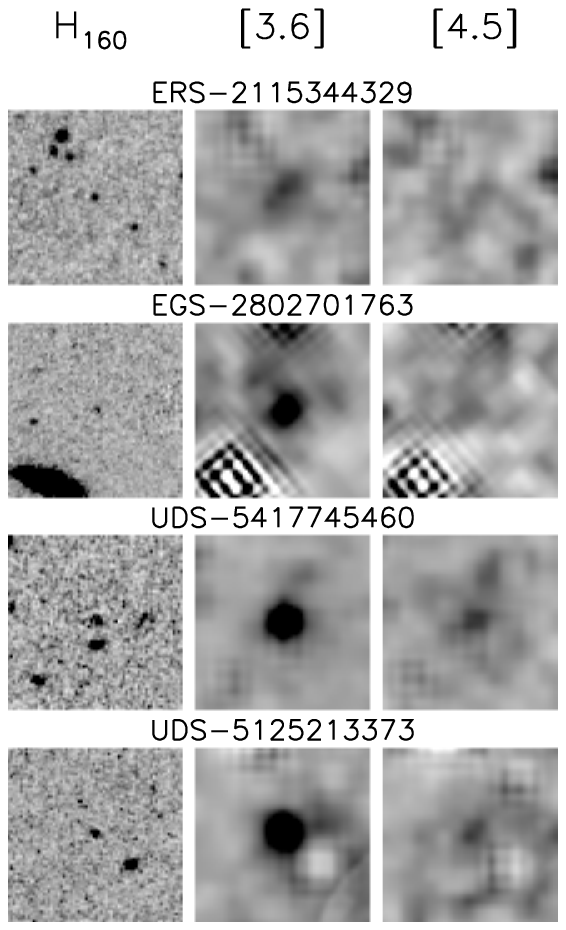}}
\caption{\textit{HST} $H_{160}$, \textit{Spitzer}/IRAC 3.6$\,\micron$, and 4.5$\,\micron$ band postage stamp negative images (8\farcs4 $\times$ 8\farcs4) of our sample of  galaxies at $z\sim6.0$ in the CANDELS fields with extremely blue $[3.6]-[4.5]$ IRAC colors (satisfying Eq. \ref{eq:blue}) and $I_{814}-J_{125}<2.3$  (see \S\ref{sec:final_sample}). The IRAC postage stamps have been cleaned for contamination from neighboring sources (see \S\ref{sec:obs_irac}). Properties of the sources are listed in Table \ref{tab:prop}. }
\label{fig:stampsz6}
\end{figure}

\begin{figure}
\centering
\includegraphics[width=0.88\columnwidth,trim=90mm 50mm 10mm 5mm] {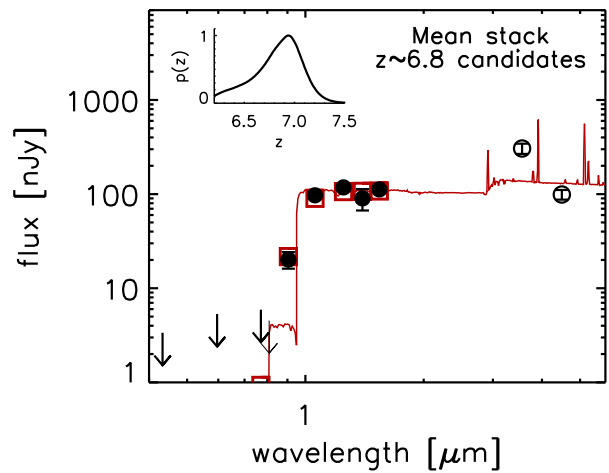} 
\caption{ Template fit to the stacked broadband observations of blue $[3.6]-[4.5]$ galaxies (Eq. \ref{eq:blue}) that also satisfy $I_{814}-J_{125}>2.3$ (see \S\ref{sec:final_sample}). Fluxes and upper limits (black points and thick arrows) show the mean \textit{HST} photometry (error bars obtained from bootstrapping). We do not include the stacked $I_{814}$, 3.6$\,\micron$ and 4.5$\,\micron$ band flux measurements (indicated by the thin arrow and open black points)  in this analysis in order to avoid biasing our photometric redshift due to our use of the $I_{814}-J_{125}$ and $[3.6]-[4.5]$ colors in selecting the sources. 
The inset panel shows the probability distribution on the mean redshift for our sample. This distribution has a mean value of $z_{\rm phot}=6.81^{+0.25}_{-0.28}$, consistent with the desired redshift range for our selection. 
} 
\label{fig:stack_z6.8}
\end{figure}

Another effect that can change the predicted $[3.6]-[4.5]$ color distribution is a probable evolution of the emission line EWs between $z\sim3$ and $z\sim6$, e.g., because we are observing increasingly younger generations of galaxies. Assuming all line EWs follow the evolution of the H$\alpha$ EW derived by \citet{Fumagalli2012} for star-forming galaxies in the redshift range $z\sim0-2$, the predicted spread in $[3.6]-[4.5]$ color would increase, while the median predicted $[3.6]-[4.5]$ color would be bluer by $\Delta ([3.6]-[4.5])=0.23$.  

From the above comparison we conclude that the blue $[3.6]-[4.5]$ sources at $z\sim6.0$ (indicated with the blue histogram in the top panel of Figure \ref{fig:hist}) can be explained by the high [\ion{O}{3}]/H$\beta$ values observed in $z\sim3$ LBGs and a blue continuum $[3.6]-[4.5]$ color as seen in young galaxies with low dust content and low metallicity, possibly in combination with an evolving EW strength of nebular emission lines as a function of redshift.

\section{A fiducial sample of $\lowercase{z}\sim6.8$ emission line galaxies}
\label{sec:final_sample}
In this section we will present a strategy for the efficient use of \textit{HST}+IRAC information to select galaxies over the redshift range $z\sim6.6-6.9$.
In \S\ref{sec:z6.8} we showed that sources at $z\sim6.6-6.9$ have very blue $[3.6]-[4.5]$ colors and that we can use this information to significantly reduce the uncertainty on the redshift determination. However, we also found that a small number of galaxies at $z<6.6$ show extremely blue colors as well and we cannot distinguish these sources from galaxies in the redshift range $z\sim6.6-6.9$ on the basis of the $[3.6]-[4.5]$ IRAC colors alone.  

\subsection{Selection of our fiducial $\lowercase{z}\sim6.8$ sample}

In order to much more effectively isolate galaxies in the narrow redshift range $z\sim6.6-6.9$, we require an additional selection criterion besides our $[3.6]-[4.5]$ color cut (Eq. \ref{eq:blue}). For our fiducial sample we require that sources show a significant break across the $I_{814}$ and $J_{125}$ bands: 
\begin{equation}
\label{eq:I-J}
I_{814}-J_{125}>2.3
\end{equation}
If $I_{814}$ is undetected we use the 1$\sigma$ upper limit to compute the color.  
This dropout criterion should effectively exclude the $z<6.5$ contaminating galaxies from our target $z\sim6.8$ selection (see Figure \ref{fig:selection}). We will separately select $z\sim6$ ultra-blue galaxies using a $I_{814}-J_{125}<2.3$ criterion.

We apply the criteria given by Eq. \ref{eq:blue} and Eq. \ref{eq:I-J} to our photometric catalogs of sources in all five CANDELS fields \citep{Bouwens2014} and find 20 sources in our fiducial $z\sim6.8$ sample and 8 sources that are likely at lower redshift. We summarize the properties of the fiducial sample and the sources that satisfy Eq. \ref{eq:blue} but not Eq. \ref{eq:I-J} in Table \ref{tab:prop}, and we show postage stamps of the sources in Figures \ref{fig:stamps} and \ref{fig:stampsz6}. The typical width of the 68\% redshift confidence intervals for the sources in our fiducial sample as given by P($z_{\rm phot,\textit{HST}+IRAC}$) is $\Delta z=0.2$.

\subsection{Ascertaining the mean redshift and contamination fraction of the $\lowercase{z}\sim6.8$ sample }
\label{sec:checks} 
 
To test the robustness of our selection we stacked the sources from GOODS-N/S selected by Eq. \ref{eq:blue} and \ref{eq:I-J} and we show the resulting spectral energy distibution (SED) in Figure \ref{fig:stack_z6.8}. We estimate the mean redshift for our selected sources from the stacked photometry.  We emphasize that we do not use our flux measurements in the $I_{814}$, 3.6$\,\micron$ and 4.5$\,\micron$ bands, due to flux measurements in these bands playing an important role in the selection of the sources themselves; this should ensure that our derived redshift is not significantly biased by the selection process itself. We find a photometric redshift of $z_{\rm phot}=6.81^{+0.25}_{-0.28}$, consistent with the redshift range $z\sim6.6-6.9$. 
Since the measured flux in the $J_{125}$ band was also used in the selection of individual sources, it could have a minor effect on the estimated redshift for the stacked photometry.  Excluding the $J_{125}$-band flux measurements in deriving the best-fit redshift, gives a photometric redshift of $z_{\rm phot}=6.77\pm0.31$. 

As a test of the robustness of our blue IRAC selection criterion (Eq. \ref{eq:blue}) against scatter in the $[3.6]-[4.5]$ color due to the uncertainties in the 3.6$\,\micron$ and 4.5$\,\micron$ flux measurements, we simulated the photometric scatter assuming an intrinsic IRAC color $[3.6]-[4.5]=0$ for all sources from the five CANDELS fields in our $z\sim5-8$ base sample.  We simulated the observed colors by adding noise to the 3.6$\,\micron$ and 4.5$\,\micron$-band fluxes, that match the measured flux uncertainties (1000$\times$ per source). From this simulation we conclude that less than 0.1 source with an intrinsic IRAC color $[3.6]-[4.5]=0$ has scattered into our fiducial selection over all five CANDELS fields.

The only significant source of interlopers to our fiducial sample of $z\sim6.8$ galaxies would seem to arise from galaxies at $z<6.6$. To quantify this interloper fraction, we first estimate the fraction of ultra-blue $[3.6]-[4.5]$ sources at $z<6.6$ from Figure \ref{fig:hist}, which is 6\%. 
From the bottom panel of Figure \ref{fig:emlines_selection} (see also Table \ref{tab:prop}) we find that sources in our fiducial sample have estimated redshifts as far as $\Delta z\sim0.3$ away from our desired redshift range $z\sim6.6-6.9$. 
We therefore assume that $z\sim6.3-6.6$ galaxies cannot be completely removed from our fiducial sample using our $I_{814}-J_{125}<2.3$ color criterion (Eq. \ref{eq:I-J}) and could potentially be contaminating our $z\sim6.8$ sample.  
Multiplying those sources from our base sample of galaxies from all five CANDELS fields (\S\ref{sec:selection}) with $z_{\rm phot,\textit{HST}}\sim6.3-6.6$ by the 6\% fraction with ultra-blue IRAC colors,  we estimate  that $\sim2.0$ sources could have scattered into our fiducial $z\sim6.8$ sample from $z<6.6$. We therefore conservatively estimate that $\sim90$\% of the 20 sources with extreme IRAC colors and $I_{814}-J_{125}$ colors $>2.3$ (Table \ref{tab:prop}) lie in the redshift range $z\sim6.6-6.9$. 

Furthermore, comparing the 20 galaxies from our $z\sim6.8$, IRAC ultra-blue sample (90\% of which we estimated to lie in this redshift range: see previous paragraph) with the 35 galaxies estimated to lie in the redshift range $z\sim6.6-6.9$ from the redshifts of our $z\sim5-8$ base sample, we estimate that $\sim$50\% of all sources at $z\sim6.8$ exhibit ultra-blue colors (vs. 6\% at $z\sim6$). This makes IRAC ultra-blue sources roughly $\sim8$ times more common at $z\sim6.8$ than at $z<6.6$. This is useful to establish, since it indicates that ultra-blue IRAC criteria - such as we propose - can potentially improve the purity of $z\sim6.6-6.9$ selections by up to a factor of 8 over what one would manage using \textit{HST} and ground-based observations alone.

\subsection{Quantifying the rest-frame EWs of [\ion{O}{3}]+H$\beta$ in our $\lowercase{z}\sim6.8$ IRAC ultra-blue sample}
\label{sec:EW}

Using the assumption that the 4.5$\,\micron$ band is free of emission line contamination at  $z\sim6.6-6.9$ while the  3.6$\,\micron$ band is contaminated by [\ion{O}{3}] and H$\beta$, we can make a prediction of the [\ion{O}{3}] strength and estimate the fraction of  high-EW [\ion{O}{3}] emitters in the high redshift galaxy population. The median $[3.6]-[4.5]$ color of our $z\sim6.8$ sample is $-1.2\pm0.3$, indicating a typical rest-frame EW$_0$([\ion{O}{3}]+H$\beta)>1000${$\,$\AA} \citep[see][]{Smit2014}.

In order to obtain a more accurate measurement of the [\ion{O}{3}] EW, we calculated the difference between the contaminated 3.6$\,\micron$ flux and an estimate of the rest-frame optical continuum flux. The continuum flux was estimated  by fitting the SEDs of the galaxies with stellar population templates using the Fitting and Assessment of Synthetic Templates (FAST) code \citep{Kriek2009}. In the fitting procedure we used stellar population templates by \citet{Bruzual2003} and constant star formation histories with ages between 30 Myr and the age of the Universe at $z=6.6$. We assumed a \citet{Salpeter} IMF with limits $0.1-100 M_\odot$ and a \citet{Calzetti2000} dust law. We considered dust contents between $A_V=0-1.5$ and subsolar metallicities between 0.2 and 0.4$Z_\odot$. 
We only considered the \textit{HST} and the 4.5$\,\micron$ IRAC photometry in deriving our best-fit model, while excluding the measured 3.6$\,\micron$ fluxes (due to their being impacted by the [\ion{O}{3}]+H$\beta$ lines). We used the best fit templates from our fitting procedure to obtain an estimate of the 3.6$\,\micron$ continuum flux and from this we derived the rest-frame EW of the combined [\ion{O}{3}]$\lambda$4959,5007{$\,$\AA}+H$\beta$ lines.

\begin{figure}
\centering
\includegraphics[width=0.88\columnwidth,trim=100mm 50mm 10mm 10mm] {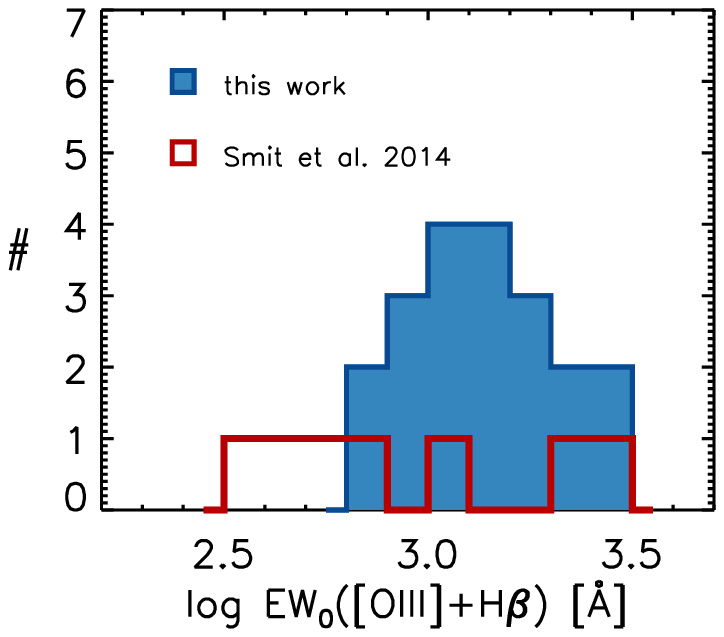} 
\caption{ The rest-frame EW distribution of [\ion{O}{3}]+H$\beta$ (blue filled histogram), estimated from the sources in our $z\sim6.8$ sample (see \S\ref{sec:EW}).
 For reference we show the EWs of the sources from \citet{Smit2014} that were selected on their photometric redshift being in the range $z\sim6.6-7.0$ (red histogram). 
 In \S\ref{sec:checks} we calculated that our fiducial sample roughly selects the $\sim50$\% strongest line emitters at $z\sim6.8$. The observed median of the distribution is 1375{$\,$\AA} rest-frame EW, but correcting for the bias in the measurement due to scatter in the $[3.6]-[4.5]$ color (for details see \S\ref{sec:EW}) we estimate a median $\rm EW_0($[\ion{O}{3}]$\rm +H\beta)$ of 1085{$\,$\AA}.  
 The excellent agreement between our sample and the 50\% most extreme sources from \citet{Smit2014} provides further evidence that high-EW nebular emission lines are indeed ubiquitous at high redshift.
} 
\label{fig:histEW}
\end{figure}

Our estimates of rest-frame EW$_0$([\ion{O}{3}]+H$\beta$) are listed in table \ref{tab:prop} and range from $\sim900${$\,$\AA} to $>2000${$\,$\AA} with a median of 1375$\,$\AA.  Though the uncertainties on the EWs are too large for the $>2000${$\,$\AA} EW measurements to be secure, we estimate rest-frame EWs as high as $\gtrsim1000${$\,$\AA} in the majority of our selected sources. Nevertheless, we note that the median EW we derive here will be biased towards high values because we are using the same $[3.6]-[4.5]$ color measurements to derive the EWs as we used for the selection. 
We estimate this bias by investigating the selection process in a distribution of galaxies with an intrinsic IRAC color $[3.6]-[4.5]=0$ and simulated the noise in the 3.6$\,\micron$ and 4.5$\,\micron$ bands based on the measured flux uncertainties. We selected 50\% of the sources with the bluest simulated $[3.6]-[4.5]$ colors  (in agreement with our estimates in \S\ref{sec:checks}) and found these sources have a median $[3.6]-[4.5]=-0.15$. Using this color bias and eq. 1 and 2 from \citet{Smit2014}  and assuming a flat continuum in $f_\nu$ for these sources, we estimate an observed bias of $\sim290${$\,$\AA} on the measured [\ion{O}{3}]+H$\beta$ EW of 1375$\,$\AA due to the noise in the IRAC bands. This suggests that the median EW of our selected sources is really 1085{$\,$\AA} in the noise-free case.

We show the predicted $\rm EW_0($[\ion{O}{3}]$\rm +H\beta)$ distribution of our $z\sim6.8$ sample in Figure \ref{fig:histEW}. We compare this distribution with the EW distribution from the seven lensed sources selected in the redshift range $z\sim6.6-7.0$ reported by \citet{Smit2014} and comparable intrinsic UV luminosities. We use the same procedure to derive EWs for these sources as described in the paragraph above. For one source we use the 1$\sigma$ upper limit due to the high uncertainties in the IRAC photometry. The comparison between the two distributions confirms that our suggested critera for selecting $z\sim6.8$ sources rougly selects the $\sim50$\% most extreme line emitters at that redshift.

The median UV luminosity of our sample (computed from the $H_{160}$-band fluxes and assuming $z=6.75$) is $M_{\rm UV}=-20.66$, roughly comparable to the $M^\ast_{\rm UV}=-20.87\pm0.26$ from the $z\sim7$ luminosity function derived by \citet{Bouwens2014}. Combining this information with our calculation in \S\ref{sec:checks} we argue that roughly $\sim50$\% of the $M^\ast_{\rm UV}$ galaxy population at $z\sim7$ produces extreme nebular emission lines (EW$_0$([\ion{O}{3}]+H$\beta)\gtrsim1000${$\,$\AA}) in the rest-frame optical.

It is interesting to compare these derived EWs to the observed [\ion{O}{3}]$\rm +H\beta$ EW distribution at $z\sim3$ reported by \citet{Schenker2013b} and \citet{Holden2014}. 
In order to match our $z\sim6.8$ sample, we use only the 50\% most extreme emitters from the combined \citet{Schenker2013b} and \citet{Holden2014} samples and calculate a median $\rm EW_0($[\ion{O}{3}]$\rm +H\beta)=390${$\,$\AA} (with a median redshift $z\sim3.5$). Comparing this number to the mean EW at $z\sim6.8$ and assuming the evolution of nebular emission line EWs scales as $\sim(1+z)^n$ \citep[see][]{Fumagalli2012}, we derive $\rm EW_0($[\ion{O}{3}]$\rm +H\beta)\propto(1+z)^{1.9\pm0.36}$. 
This slope is consistent with the slope derived  by \citet{Fumagalli2012} for H$\alpha$ EWs over the redshift range $z\sim0-2$ ($n\sim1.8$) but somewhat steeper than the slope derived by \citet{Labbe2013}, who found $n=1.2\pm0.25$ over the redshift range $z\sim1-8$.

\section{Summary and Discussion}
\label{sec:Summary}

In this paper we explore the use of IRAC colors to select star-forming galaxies in the narrow redshift range $z\sim6.6-6.9$. Sources in this redshift range are expected to be very blue due to the boosted flux in the 3.6$\,\micron$ band from high-EW [\ion{O}{3}] emission, while the 4.5$\,\micron$ band is free from contamination of strong nebular emission lines. This suggests a blue IRAC criterion may be appropriate for selecting galaxies in the redshift range $z\sim6.6-6.9$. 

In evaluating the suitability of such a selection criterion we analyze a large sample of Lyman Break galaxies in GOODS-N and GOODS-S with relatively high $S/N$ \textit{Spitzer}/IRAC coverage from the \textit{Spitzer} GOODS, ERS, and S-CANDELS programs (Ashby et al., submitted). We find that the majority of candidates with extremely blue $[3.6]-[4.5]$ colors are consistent with $z\sim6.8$. In addition to the $z\sim6.6-6.9$ sources, we also find a small number of sources with extreme $[3.6]-[4.5]$ colors more likely to be at $z\sim6.0$. The blue $[3.6]-[4.5]$ galaxies at $z\sim6.0$ can be explained by high [\ion{O}{3}]/H$\beta$ ratios, such as found in $z\sim3$ LBGs \citep{Schenker2013b,Holden2014}, lensed galaxies at $z\sim1-2$ \citep{Brammer2012b,vanderwel2013}, and in Green Pea galaxies at $z\sim0$ \citep{Amorin2012,Jaskot2013} in combination with blue colors for the optical continuum. 

To obtain a clear selection in the redshift range $z\sim6.6-6.9$, we suggest the use of ultra-blue IRAC colors combined with an $I_{814}-J_{125}>2.3$ dropout criterion. Based on our analysis in section \S\ref{sec:final_sample}, we estimate that at least 90\% of the sources selected by these criteria lie in the redshift range $z\sim6.6 - 6.9$. We systematically apply such criteria to our source catalogs from the five CANDELS fields (720 arcmin$^2$) and find 20 sources ($\sim0.03$ arcmin$^{-2}$). A comparison with the total number of galaxies from our catalogs in this redshift range suggests that we select the $\sim50$\% bluest IRAC sources at $z\sim6.8$ at a typical UV-luminosity $M_{\rm UV} \sim M^\ast_{\rm UV, z=7}$. 

 We estimate a median uncertainty on the redshift estimates for our fiducial sources of $\Delta z=0.2$ (68\% confidence). Such redshift uncertainties are significantly smaller than one finds for these objects without the inclusion of the IRAC fluxes. For example, we find a median 68\% confidence interval on $z_{\rm phot,\textit{HST}}$ of $\Delta z=0.6$ and $\Delta z=1.7$, respectively, for sources in GOODS-N/S and COSMOS/EGS/UDS. Among other uses, such tight constraints on the redshifts are necessary for efficient observations with ALMA.  Our constraints on the redshift of these sources means that we should typically only require two ALMA tunings to successfully observe [\ion{C}{2}]$\lambda$157.7$\,\micron$ in band 6.

Using our fiducial sample of $z\sim6.8$ sources with ultra-blue $[3.6]-[4.5]$ colors, we estimate the strength of [\ion{O}{3}]$\lambda$4959,5007{$\,$\AA}+H$\beta$ from the contaminated 3.6$\,\micron$ flux. We find that the majority of the sources in our sample show EWs as high as EW$_0$([\ion{O}{3}]+H$\beta)\gtrsim1000${$\,$\AA}, in excellent agreement with the $\sim50$\% most extreme sources from \citet{Smit2014}.

Given the recent study by \citet{Stark2014a}, who found evidence for strong [\ion{O}{3}]$\lambda\lambda$4959,5007{$\,$\AA} emission in sources with high-EW \ion{C}{3}]$\lambda$1908{$\,$\AA} lines at $z\sim2$, it seems reasonable to suppose that our strong [\ion{O}{3}] emitters also exhibit relatively strong \ion{C}{3}] lines, and therefore our sources are excellent targets for follow-up studies with near-IR spectroscopy \citep[e.g.][]{Stark2014b}.

Finally, with the future launch of the \textit{JWST}, we will be able to target these extreme line-emitter galaxies with \textit{JWST}'s near infrared spectrograph (NIRSpec). The  [\ion{O}{3}]$\lambda$5007{$\,$\AA} line in a typical galaxy from our sample should be detected at 10$\sigma$ in the \textit{JWST}/NIRSpec $R=100$ mode with a mere 60 second exposure.

\acknowledgments
We thank Jarle Brinchmann, Rob Crain, Paul van der Werf and Tim van Kempen for useful discussions. We are grateful to Brad Holden for providing us with the complete catalog from \citet{Holden2014}. We acknowledge support from ERC grant HIGHZ \#227749, and an NWO vrij competitie grant.

\bibliographystyle{plainnat}

\begin{table*}
\scalebox{0.95}{
\begin{threeparttable}
\centering
\caption{Properties of extremely blue $[3.6]-[4.5]$ galaxies in our samples}
\label{tab:prop}
\tabcolsep=0.15cm
\begin{tabular}{ccccccccc}
\hline
\hline
{ID}  & {RA} & {Dec}  & {$m_{H_{160}}$} & {$I_{814}-J_{125}$} & {$[3.6]-[4.5]$} & {$z_{\rm phot,\textit{HST}}^c$} & {$z_{\rm phot,\textit{HST}+IRAC}^c $}  & {EW([\ion{O}{3}]+H$\beta$)[$\,$\AA]}  \\ 
\hline
\multicolumn{8}{c}{Fiducial $z\sim6.8$ sample$^a$} \\
\hline
GND-7035815571 & $12{:}37{:}03.586$ & $+62{:}15{:}57.18$ & $26.3 \pm 0.1$ & $ > 3.2$ & $-1.0 \pm 0.3$ & $7.26^{+0.33}_{-0.34} $ & $6.81^{+0.07}_{-0.07} $ & $ 1374 \pm 415$  \\
GND-7025017221 & $12{:}37{:}02.500$ & $+62{:}17{:}22.17$ & $26.8 \pm 0.1$ & $ > 2.8$ & $-1.2 \pm 0.3$ & $7.25^{+0.43}_{-0.45} $ & $6.76^{+0.06}_{-0.06} $ & $ 1779 \pm 486$  \\
GND-6427518385 & $12{:}36{:}42.753$ & $+62{:}18{:}38.55$ & $27.0 \pm 0.2$ & $2.5 \pm 0.9$ & $-1.5 \pm 0.4$ & $6.52^{+0.37}_{-0.37} $ & $6.71^{+0.08}_{-0.06} $ & $ 2591 \pm 1095$  \\
GND-6372717115 & $12{:}36{:}37.279$ & $+62{:}17{:}11.59$ & $26.3 \pm 0.1$ & $ > 3.2$ & $-0.9 \pm 0.3$ & $6.75^{+0.27}_{-0.26} $ & $6.70^{+0.13}_{-0.12} $ & $ 1193 \pm 485$  \\
GSD-2504846559 & $03{:}32{:}50.481$ & $-27{:}46{:}55.95$ & $25.8 \pm 0.1$ & $2.3 \pm 0.9$ & $-0.9 \pm 0.2$ & $6.83^{+0.32}_{-0.34} $ & $6.78^{+0.08}_{-0.08} $ & $ 962 \pm 293$  \\
GSD-2254450533 & $03{:}32{:}25.443$ & $-27{:}50{:}53.36$ & $26.3 \pm 0.1$ & $ > 3.3$ & $-1.3 \pm 0.3$ & $6.96^{+0.26}_{-0.26} $ & $6.78^{+0.06}_{-0.06} $ & $ 1471 \pm 432$  \\
GSD-2237749136 & $03{:}32{:}23.778$ & $-27{:}49{:}13.64$ & $26.6 \pm 0.1$ & $ > 3.0$ & $-1.3 \pm 0.4$ & $6.90^{+0.20}_{-0.19} $ & $6.74^{+0.09}_{-0.09} $ & $ 1800 \pm 704$  \\
GSD-2252146266 & $03{:}32{:}25.216$ & $-27{:}46{:}26.69$ & $26.9 \pm 0.1$ & $ > 3.1$ & $-1.6 \pm 0.5$ & $6.94^{+0.23}_{-0.23} $ & $6.75^{+0.06}_{-0.06} $ & $ 2424 \pm 1083$  \\
COS-3731073631 & $10{:}00{:}37.310$ & $+02{:}27{:}36.31$ & $26.0 \pm 0.1$ & $ > 2.3$ & $-1.5 \pm 0.5$ & $7.43^{+1.07}_{-1.23} $ & $6.68^{+0.14}_{-0.14} $ & $ 1686 \pm 676$  \\
COS-2987030247$^{d}$ & $10{:}00{:}29.870$ & $+02{:}13{:}02.47$ & $24.8 \pm 0.1$ & $2.7 \pm 0.8$ & $-1.0 \pm 0.1$ & $6.99^{+1.16}_{-1.25} $ & $6.66^{+0.14}_{-0.14} $ & $ 1128 \pm 166$  \\
COS-1318939512$^{e}$ & $10{:}00{:}13.189$ & $+02{:}23{:}9.512$ & $25.0 \pm 0.1^{e}$ & $2.8 \pm 1.0$ & $-1.5 \pm 0.4$ & $7.15^{+1.07}_{-0.99} $ & $6.75^{+0.09}_{-0.08} $ & $ 786 \pm 301$  \\
COS-3018555981$^{f}$ & $10{:}00{:}30.185$ & $+02{:}15{:}59.81$ & $24.9 \pm 0.1$ & $ > 3.3$ & $-1.2 \pm 0.1$ & $7.76^{+0.79}_{-0.82} $ & $6.76^{+0.07}_{-0.07} $ & $ 1424 \pm 143$  \\
EGS-4434164969 & $14{:}19{:}44.341$ & $+52{:}56{:}49.69$ & $26.3 \pm 0.1$ & $ > 2.6$ & $-1.3 \pm 0.5$ & $7.08^{+1.41}_{-1.82} $ & $6.62^{+0.19}_{-0.20} $ & $ 1321 \pm 619$  \\
EGS-5711424617 & $14{:}19{:}57.114$ & $+52{:}52{:}46.17$ & $25.1 \pm 0.1$ & $2.4 \pm 0.2$ & $-0.8 \pm 0.1$ & $6.33^{+0.22}_{-0.21} $ & $6.47^{+0.10}_{-0.10} $ & $ 927 \pm 149$  \\
EGS-3506853076$^{g}$ & $14{:}19{:}35.068$ & $+52{:}55{:}30.76$ & $26.2 \pm 0.1$ & $2.9 \pm 0.9$ & $-0.9 \pm 0.2$ & $7.18^{+0.94}_{-0.82} $ & $6.62^{+0.16}_{-0.16} $ & $ 1084 \pm 298$  \\
EGS-1952445714$^{h}$ & $14{:}19{:}19.524$ & $+52{:}44{:}57.14$ & $25.3 \pm 0.1$ & $ > 3.9$ & $-0.7 \pm 0.1$ & $7.52^{+0.66}_{-0.66} $ & $6.75^{+0.11}_{-0.11} $ & $ 768 \pm 151$  \\
EGS-1350184593 & $14{:}19{:}13.501$ & $+52{:}48{:}45.93$ & $26.5 \pm 0.1$ & $ > 3.2$ & $-1.6 \pm 0.6$ & $7.56^{+0.71}_{-0.74} $ & $6.72^{+0.09}_{-0.08} $ & $ 2391 \pm 1196$  \\
EGS-2713432218 & $14{:}19{:}27.134$ & $+52{:}53{:}22.18$ & $26.1 \pm 0.1$ & $ > 3.1$ & $-0.8 \pm 0.2$ & $7.56^{+0.66}_{-0.68} $ & $6.71^{+0.13}_{-0.14} $ & $ 1048 \pm 242$  \\
UDS-0089122444 & $02{:}17{:}00.891$ & $-05{:}12{:}24.44$ & $26.5 \pm 0.2$ & $ > 2.7$ & $-1.5 \pm 0.6$ & $7.34^{+0.83}_{-0.87} $ & $6.70^{+0.11}_{-0.10} $ & $ 2620 \pm 1350$  \\
UDS-5754844803$^i$ & $02{:}17{:}57.548$ & $-05{:}08{:}44.80$ & $24.8 \pm 0.2$ & $2.5 \pm 1.1$ & $-0.9 \pm 0.1$ & $7.05^{+1.08}_{-1.28} $ & $6.61^{+0.19}_{-0.17} $ & $ 915 \pm 95$  \\

\hline
\multicolumn{8}{c}{$z\sim6.0$ sample$^{b}$} \\
\hline
GND-6322718286 & $12{:}36{:}32.273$ & $+62{:}18{:}28.67$ & $26.2 \pm 0.1$ & $2.0 \pm 0.4$ & $-1.1 \pm 0.3$ & $7.01^{+0.23}_{-0.24} $ & $6.75^{+0.08\, j}_{-0.09} $ & $-$  \\
GNW-6390808452 & $12{:}36{:}39.080$ & $+62{:}08{:}45.29$ & $26.1 \pm 0.2$ & $2.1 \pm 0.8$ & $-0.7 \pm 0.1$ & $5.87^{+0.33}_{-0.33} $ & $4.84^{+1.50\, k}_{-0.03} $ & $-$  \\
GSD-2234402156 & $03{:}32{:}23.449$ & $-27{:}50{:}21.56$ & $26.5 \pm 0.1$ & $1.5 \pm 0.3$ & $-1.0 \pm 0.3$ & $5.97^{+0.22}_{-0.22} $ & $6.09^{+0.13\, k}_{-0.12} $ & $-$  \\
GSW-2573853217 & $03{:}32{:}57.381$ & $-27{:}53{:}21.78$ & $26.2 \pm 0.2$ & $ > 1.7$ & $-0.8 \pm 0.2$ & $6.27^{+0.25}_{-0.26} $ & $6.46^{+0.13\, k}_{-0.13} $ & $-$ \\
ERS-2115344329 & $03{:}32{:}11.539$ & $-27{:}44{:}32.99$ & $26.9 \pm 0.1$ & $1.1 \pm 0.2$ & $-1.0 \pm 0.4$ & $5.84^{+0.25}_{-0.30} $ & $6.03^{+0.12\, k}_{-0.21} $ & $-$ \\
EGS-2802701763 & $14{:}20{:}28.027$ & $+53{:}00{:}17.63$ & $26.5 \pm 0.2$ & $1.1 \pm 0.3$ & $-1.4 \pm 0.5$ & $5.48^{+0.54}_{-0.84} $ & $5.94^{+0.14\, k}_{-0.24} $ & $-$  \\
UDS-5417745460 & $02{:}17{:}54.177$ & $-05{:}14{:}54.60$ & $25.5 \pm 0.1$ & $2.2 \pm 0.4$ & $-0.7 \pm 0.2$ & $6.32^{+0.29}_{-1.23} $ & $6.45^{+0.18\, k}_{-0.19} $ & $-$ \\
UDS-5125213373 & $02{:}17{:}51.252$ & $-05{:}11{:}33.73$ & $25.6 \pm 0.1$ & $1.8 \pm 0.3$ & $-0.9 \pm 0.3$ & $6.01^{+0.34}_{-1.36} $ & $6.25^{+0.14\, k}_{-0.15} $ & $-$  \\

\hline
\hline
\end{tabular}
\begin{tablenotes}
\item $^a${In addition to our blue IRAC criterion (see Eq. \ref{eq:blue}) we require $I_{814}-J_{125}>2.3$ for our fiducial $6.6-6.9$ sample. If $I_{814}$ is undetected we use the 1$\sigma$ upper limit to compute the color. }
\item $^b${Here, we list sources with $I_{814}-J_{125}<2.3$, as well as blue IRAC colors (satifying Eq. \ref{eq:blue}). Low $S/N$ sources cannot unambiguously be selected in the $z\sim6.0$ sample, i.e., if they satisfy $I_{814}-J_{125}<2.3\,\wedge\,S/N(I_{814})<1$. However, this is only the case for one source in our sample, i.e. GSW-2573853217. The photometric redshift estimate of this source indicates it likely belongs in the $z\sim6.0$ sample.    }
\item $^c${Error bars indicate the 68\% confidence interval.}
\item $^d${In the stacked ground-based optical image (inverse variance weighted) this source is detected at $7.9\pm1.5$ nJy. However, this flux appears to derive from a foreground source, close to our object of interest but distinctly separated in the HST optical images. }
\item $^e${This source is detected at the edge of the \textit{HST}/WFC3 field of view. We have verified that this source is also detected in the ground-based photometry from the UltraVISTA survey \citep[see][]{Ilbert2013}. However, the total magnitude for this source measured from ground-based data appears to be somewhat fainter than we measure for HST, suggesting that our HST stack might be affected by some non-gaussian noise.  }
\item $^f${This source was independently discovered by \citet{Tilvi2013} and \citet{Bowler2014}, who estimated a photometric redshift of $7.24^{+0.38}_{-0.25}$ and $6.77^{+0.14}_{-0.19}$ respectively.}
\item $^g${This source is only marginally resolved; while its spatial profile and SED are much more consistent with its being a $z\sim6.8$ galaxy, we cannot completely exclude the possibility it is a low-mass star.}
\item $^h${Though the size of this source is consistent with its being a low-mass star, the SED of this source is better fit by a high-redshift galaxy than a stellar tempate.}
\item $^i${This source, better known as 'Himiko' \citep{Ouchi2009,Ouchi2013}, has a spectroscopic redshift at $z_{Ly\alpha}=6.59$, consistent with our photometric redshift estimate within the 68\% probability window.}
\item $^j${This source is weakly detected in $I_{814}$ with $I_{814}-J_{125}=2.0\pm0.4$ and therefore included in the $z\sim6$ sample. However, due to a $<1\sigma$ detection in the $z_{850}$ band, our estimated photometric redshift indicates a $z\sim6.75$ solution.}
\item $^k${The typical uncertainty in $z_{\rm phot,\textit{HST}+IRAC}$ for this $z\sim6$ sample is very small. The ultra-blue IRAC colors of these galaxies are preferentially fit by the template with the most extreme [\ion{O}{3}]/H$\alpha$ ratio, which allows for little variation of the colors of the continuum emission and thereby narrows the redshift probability distribution. However, the shape of the spectral energy distributions of these galaxies (and the range of [\ion{O}{3}]/H$\alpha$ ratios at $z\sim6$) is unknown, and therefore it is likely that the width of the redshift probability distribution is underestimated for this particular sample.    }
\end{tablenotes}
\vspace{0.3cm}
\end{threeparttable}
}
\end{table*}

\end{document}